\documentclass[onecolumn,letterpaper,amsmath,amssymb,floatfix,aps,preprintnumbers,
showpacs,superscriptaddress]{revtex4}

\usepackage{verbatim}
\usepackage{graphicx}
\usepackage{bm} 
\usepackage{dcolumn}  
\usepackage{epsfig}
\usepackage{color}

\newcommand{\nn}{\nonumber}
\newcommand{\la}{\langle}
\newcommand{\ra}{\rangle}
\newcommand{\bacol}{\setlength{\arraycolsep}{0pt}}

\begin{document}

\title{Wrapping Transition and Wrapping-Mediated Interactions for Discrete Binding along an Elastic Filament: An Exact Solution}

\author{David S. \surname{Dean}}
\affiliation{Universit\'e de  Bordeaux and CNRS, Laboratoire Ondes et Mati\`ere d'Aquitaine (LOMA), 
UMR 5798, F-33400 Talence, France}
\author{Thomas~C. \surname{Hammant}}
\author{Ronald~R. \surname{Horgan}}
\affiliation{Department~of~Applied~Mathematics~and~Theoretical~Physics,
University~of~Cambridge, Centre~for~Mathematical~Sciences, Cambridge~CB3~0WA, United~Kingdom}
\author{Ali \surname{Naji}}
\thanks{Corresponding author -- email: \texttt{a.naji@ipm.ir}}
\affiliation{School of Physics, Institute for Research in
Fundamental Sciences (IPM), Tehran 19395-5531, Iran }
\affiliation{Department~of~Applied~Mathematics~and~Theoretical~Physics,
University~of~Cambridge, Centre~for~Mathematical~Sciences, Cambridge~CB3~0WA, United~Kingdom}
\author{Rudolf \surname{Podgornik}}
\affiliation{Department of Theoretical Physics, J. Stefan Institute, SI-1000 Ljubljana, Slovenia} 
\affiliation{Department of Physics, Faculty of Mathematics and Physics, University of Ljubljana, SI-1000 Ljubljana, Slovenia}
\affiliation{Department of Physics, University of Massachusetts, Amherst, MA 01003, USA}

\begin{abstract} 
The wrapping equilibria of one and two adsorbing cylinders are studied  along a semi-flexible filament (polymer) due to the interplay between elastic rigidity and short-range adhesive energy between the cylinder and the filament.  
We show that statistical mechanics of  the system can be solved exactly using a path integral formalism which gives access to the full effect of thermal fluctuations, going thus beyond the usual Gaussian approximations which take into account only the contributions from the minimal energy configuration and small fluctuations about this minimal energy solution. We obtain the free energy of the wrapping-unwrapping transition of the filament around the cylinders as well as the effective interaction between two wrapped cylinders due to thermal fluctuations of the elastic filament. 
A change of entropy due to wrapping of the filament around the adsorbing cylinders as they move closer together is identified as an additional source of interactions between them. Such entropic  wrapping effects should be distinguished from the usual entropic configuration effects in semi-flexible polymers. Our results may be applicable to the problem of adsorption of proteins as well as synthetic nano-particles on semi-flexible polymers such as DNA. 
\end{abstract}


\maketitle

\section{Introduction}

Interaction, binding and complexation to membranes and polymers, as typified by  protein adsorption to DNA  chains,  play a vital role in biology and biochemistry.  In biology, for 
example, a variety of proteins interact with DNA in cellular processes involving 
the reading and packaging of the genome \cite{Alberts}. Transcription 
factors are proteins which bind to specific DNA sequences controlling 
the flow of genetic information to messenger RNA (mRNA) which in turn 
carries this information to ribosomes for protein synthesis. DNA polymerase is 
another type of protein (enzyme) that plays a key role in the DNA replication process: 
it synthesizes a new DNA strand by reading genetic information from an intact 
DNA strand that serves as a template. RNA polymerase is a 
different kind of DNA-binding protein that binds to DNA and uses it as a 
template to synthesize the RNA. In all these examples, protein binding induces local deformation of DNA.  When multiple proteins bind to DNA they may exhibit a 
significant degree of cooperativity \cite{Sun}, which may result from the 
formation of loops  \cite{Schlief}, specific protein-protein interactions (at short separations) \cite{Lilley} or a  variable-range cooperative binding of proteins regulated by the tension along 
the DNA strand \cite{rubr:1999}.

Perhaps the most remarkable form of protein-induced deformation of DNA is found 
in the {\em chromatin fibre}. In eukaryotic cells, a long strand of DNA 
is  efficiently packed within a micron-size cell nucleus. This occurs through a 
hierarchical folding process \cite{Alberts} where at the lowest length scale, short 
segments (about 50~nm)  of DNA are wrapped around small histone proteins forming  {\em nucleosomes}: cylindrical wedge-shaped histone octamers of diameter of around 7~nm and mean height 5.5~nm. DNA-histone binding 
may be mediated through specific binding sites \cite{Schiessel_rev} as well as 
non-specific electrostatic interactions \cite{phys_rep,Kuno1,Kuno2} since both 
DNA and histones carry relatively large opposite (net) charges (on the average,  around 
188, 130 and 129 histone charges interact with the 
wrapped DNA for native chromatin, nucleosome core particle and H1-depleted 
chromatin, respectively \cite{Kimura}, while DNA itself carries 6 elementary charges per nm). 
In nucleosomes,  DNA is strongly bent  and wrapped around the core histones in nearly a 1-and-3/4 left-handed helical 
turn, which thus costs a large bending energy given that DNA has an 
effective persistence length of around 50~nm in physiological conditions 
\cite{Rief99}. 

These DNA-histone complexes are linked together and, on the next level of the 
hierarchy, fold into a rather dense structure known as the 30 nm chromatin 
fibre, which undergoes a series of higher-order foldings resulting in highly 
condensed chromosomes. Under physiological conditions (salt concentration of 
around 100mM NaCl), this fibre exhibits a diameter of about 30~nm, while at low
salt concentrations the fibre is swollen and displays a beads-on-a-string 
pattern with a diameter of around 10~nm in which core particles are widely 
separated \cite{Schiessel_rev,Woodcock_rev,vanHolde}. In these cases, the 
wrapped structure of DNA around histone proteins remains intact. However, at 
salt concentrations below 1~mM, the DNA begins to partially unwrap from the 
histone cores \cite{Yager}. The electrostatic mechanism for the salt-induced 
wrapping-unwrapping behavior of the DNA around nucleosome core particles has been discussed 
in the literature for  general classes of charged polymer-macroion complexes 
\cite{Kuno1,Kuno2,hoda1,hoda2,hoda3} and for long complex fibres of multiple 
macroions \cite{hoda4} and will not be considered in this paper.  In general, the 
interplay between the DNA's bending energy cost, the electrostatic attraction energy 
as well as other specific binding energies would be key to understanding the global
structural features of the chromatin fiber. The precise 
arrangement of nucleosome core particles in the 30~nm fibre is still intensely 
debated in the literature (see, e.g.  \cite{Schiessel_rev,Woodcock_rev, 
vanHolde, Thoma, Woodcock, Bednar, Horowitz, Worcel, Wedemann, Schiessel_EPL, 
Richmond05, Robinson2006, Langowski1, Langowski2, Diesinger2, Depken} and 
references therein).

Genome-wide experimental mappings of nucleosome occupancy in yeast point to a 
patchy landscape composed of at least partly ordered crystal-like configurations 
with a nucleosome repeat length of about 165 base-pairs \cite{Kaplan}. It appears that 
what is encoded in the sequence of DNA is not the sequential ordering of  
nucleosomes but rather the {\em nucleosome excluded} regions. In between these regions 
nucleosomes position themselves via a thermodynamic equilibrium mechanism
\cite{Arneodo1}.  The ordered configurations of nucleosomes between the excluded 
regions have been modeled as a nonuniform fluid of one-dimensional (1D) hard rods confined by two 
excluding energy barriers at the extremities \cite{Arneodo1,Arneodo2} (see Ref.  \cite{hoda4}
for a ground-state treatment in three dimensions). The interaction between the nucleosomes 
were thus assumed to be of purely hard-core  type, i.e., only  steric exclusion is taken into account. 
Understanding the details of these interactions and the adsorption-wrapping equilibria for 
protein-DNA complexes is thus of paramount importance. 

Motivated by these findings, we focus here on elastic properties and 
adsorption-desorption equilibria of a model system composed of adsorbing 
cylindrical particles and an elastic filament modeling a semi-flexible polymer chain with short-range 
adsorption interaction, assumed to be proportional to the arc-length of the 
filament touching the particle. We neglect any electrostatic 
interactions (which may be justified for charged species
at high enough salt concentration in the solution) and focus  on 
elasticity and short-range (adhesive) wrapping. Furthermore, we restrict 
ourselves to a 2D Eulerian plane model of a single elastic filament wrapped around 
one, two or many cylinders. This formulation of the model system 
allows us to introduce an {\em exact} formalism based on the Schr\"odinger 
representation for the partition function of the system and to carry out the 
calculations explicitly for the wrapping-unwrapping behavior of the elastic filament 
with one or two wrapping cylinders as well as for the effective interaction  
between two cylinders wrapped on a single filament.
Our study may thus be applicable to the problem of adsorption of proteins as well as  synthetic 
nano-particles to semi-flexible polymers such as DNA.

In the same context, the tension-mediated interaction between proteins bound to a DNA chain was
initially considered by Rudnick and Bruinsma \cite{rubr:1999} within the very same 
model  used in the present paper. The thermal fluctuations of the chain were 
however dealt with only on a {\em harmonic} level, corresponding to small Gaussian 
fluctuations around the ground state. Our analysis, based on the Schr\"odinger 
representation, goes beyond this approximation and takes into account the whole 
spectrum of chain conformational fluctuations within a rigorous mathematical 
framework. In agreement with Ref. \cite{rubr:1999}, we find that the interaction 
between two wrapped cylinders may be either attractive or repulsive, depending on 
their relative orientation along the elastic filament. The details of the 
interaction are however different from those found on the simple Gaussian level.

In another recent related work, Koslover and Spakowitz \cite{Spakowitz} studied 
the role of twist in the coupling between bound proteins. It was shown that 
twist resistance results in a more complex interaction between the cylinders 
exhibiting, e.g., damped oscillations superimposed over and counteracting the 
attractive protein-protein interaction. In this paper, confining ourselves to 
the 2D Eulerian plane, we shall neglect the twist contribution and focus 
primarily on exact solutions for the case where the polymer is described as a
 twist-free worm-like chain.


The organization of the paper is as follows: In Section \ref{theory}, we define our model and discuss the general formalism which we shall employ 
in our study. In Section \ref{one_wa}, we apply our formalism to the problem of the unwrapping transition for a single wrapped cylinder and in 
Section \ref{unbinding}, we study the unwrapping transition of the filament from two cylinders. In Section \ref{effective_interaction}, we consider the 
problem of effective interaction mediated by the elastic filament's fluctuations between two wrapped cylinders. We summarize our results and conclude
 in Section \ref{sec:sum}. 

\section{\label{theory} Filament wrapping around cylinders: Model and formalism}

\subsection{Filament elastic energy}

\begin{figure}[t]
   \centering
  \includegraphics[width=9cm]{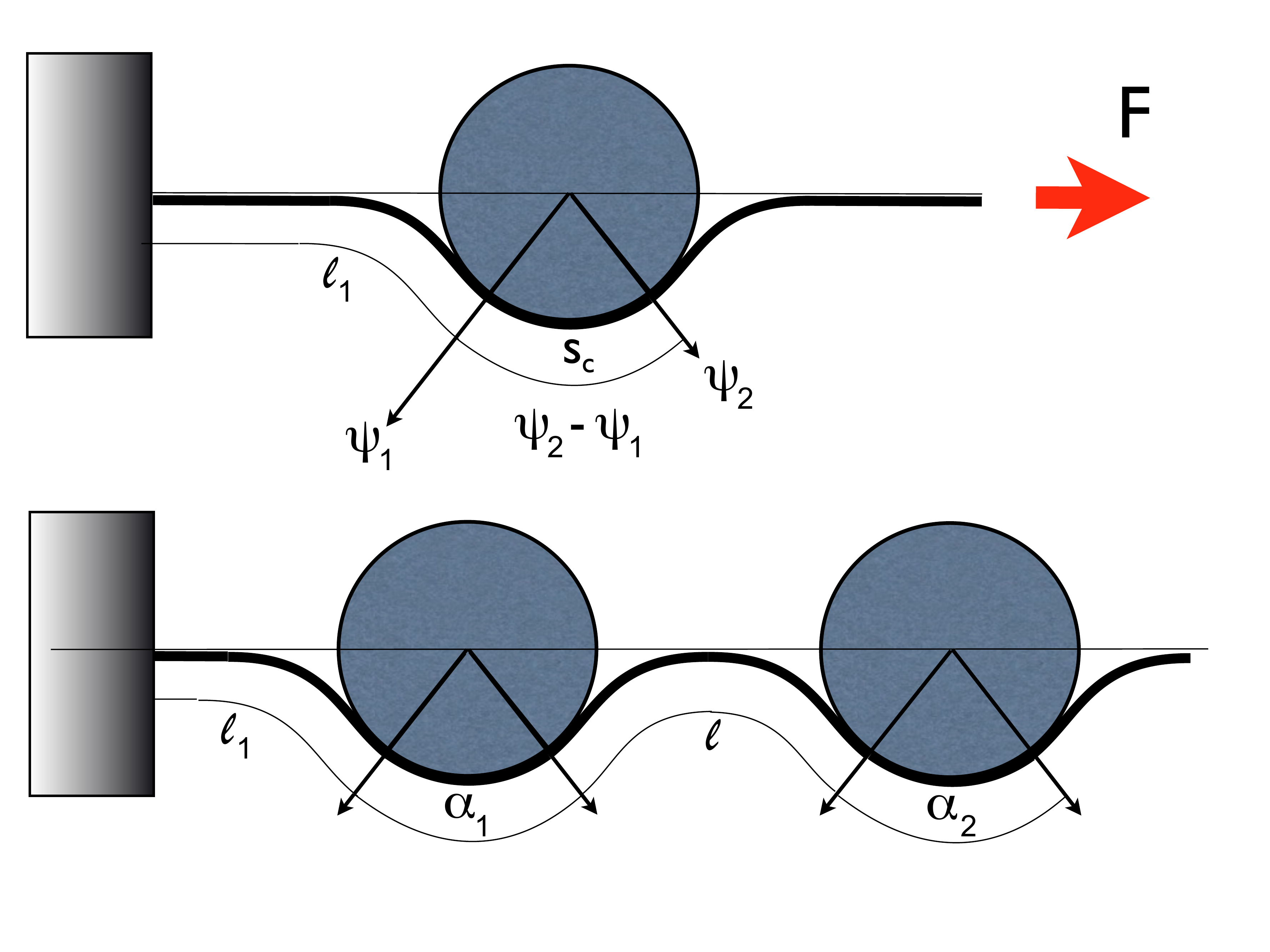}
  \caption{A schematic drawing of the geometry as well as various quantities defined in the 
text for one and two cylinders wrapped on the Eulerian filament (shown in a symmetric mode). 
Specifically, for one cylinder (top) the angle at which the filament 
first touches the cylinder (advancing along the arc-length to the cylinder) 
is $\psi_1$ and the angle at which it leaves is $\psi_2$. For two cylinders (bottom), separated by a distance $l$, the contact angles 
 $\alpha_1$  and $\alpha_2$ are as shown in the figure. }
  \label{schematic}
\end{figure}

Consider an inextensible semi-flexible filament of total arc-length $L$ 
confined to a 2D Eulerian plane with coordinates $(x,y)$. The 
tangent vector of the filament is normalized to one and can thus be 
parametrized as ${\bf t}(s) = {\dot{\bf x}}(s) = (\cos(\psi(s)),\sin(\psi(s)).$
In this parametrization the inextensibility constraint is taken into account exactly.
The bending energy of the filament is given by
\begin{equation}
\int_0^L ds\ {\kappa\over 2}\left({d^2{\bf x}\over ds^2}\right)^2 = 
\int_0^L ds\ {\kappa\over 2}\bigg({d{\psi}\over ds}\bigg)^2, 
\label{eq:bending-energy}
\end{equation}
where $\kappa$ is the filament stiffness parameter and has dimensions of energy 
times length (the so-called ``persistence length" is defined as $L_p=\kappa/(k_BT)$ for the ambient temperature $T$). We assume that one end of the filament is fixed at ${\bf x}(s=0)=0$ and the other 
end is pulled with an external tension ${\bf F}$ in the direction $x$, see Fig. \ref{schematic}. The potential energy of the 
filament is thus given by
\begin{equation}
-Fx(L)=-F\int_0^L ds\ \cos(\psi(s)). 
\end{equation}
The total energy of the filament without any interactions with objects in the plane is then given as an 
elastic energy functional $E[\psi] $ as
\begin{equation}
E(L)  = E[\psi] = \int_0^L ds  \left[{\kappa\over 2}\bigg({d{\psi}\over ds}\bigg)^2 - F \cos(\psi(s)) \right],
\label{enfunc}
\end{equation}
which is the standard Euler elastic energy expression for the elastic filament under external force.

The energy function Eq. (\ref{enfunc}) also arises in the study of parallel cylinders
adhering to elastic {\em membranes} \cite{weik,muller,mkr,muller2}. In this case, when the cylinders are parallel, and if only membrane height fluctuations in the direction normal to the cylinders are taken into account, 
the problem becomes effectively one dimensional and if one uses the Helfrich Hamiltonian for the energy of the membrane due to its height
fluctuations one can write (up to a constant term) \cite{muller2} 
\begin{equation}
E_{memb} = {\cal L}\int_0^L ds  \left[{\kappa\over 2}\bigg({d{\psi}\over ds}\bigg)^2 - \sigma \cos(\psi(s)) \right],\label{meme}
\end{equation}
where $\psi(s)$ is the tangent angle of the membrane in the direction perpendicular to the cylinders ($s$ being the coordinate in this direction), $L$ is the length of membrane in this direction and $\cal L$ the
length parallel to the cylinders. The term $\kappa$ is the membrane bending rigidity and $\sigma$ the surface tension.  
The Hamiltonian Eq. (\ref{meme}) is thus formally the same as that studied by Rudnick and Bruinsma \cite{rubr:1999} as given by Eq. (\ref{enfunc}); however there are a number of
notable and crucial differences between the two models. First, Eq. (\ref{meme}) is multiplied by a 
macroscopic quantity $\cal L$, this means that the statistical mechanics can be determined by purely 
energetic (or mean field) considerations by minimizing Eq. (\ref{meme}). In this limit one can also assume that fluctuations are small and thus fluctuations can be taken into account via an expansion about
the minimal energy configuration and treated harmonically  as was done by Rudnick and Bruinsma \cite{rubr:1999} for the polymer model. Another technical difference is that when one considers the interaction between the cylinders in the membrane model, the force is calculated as a function of the  spatial distance between the cylinders, whereas in the polymer problem it is
more natural to consider the arc-length along the polymer separating the cylinders as the relevant 
physical parameter. However within the formalism introduced here both ensembles (fixed spatial distance and fixed arc-length) can be handled on an equal footing. 
This choice of {\em ensemble} is especially important when analyzing the effect
of fluctuations. The above points thus conclude the technical motivation for our study: as the problem is effectively one dimensional, one can be in a limit where the effects of fluctuations are not merely perturbative and a full treatment of fluctuations is thus necessary. The limit where the energetic
minimization is valid in the treatment of Eq. (\ref{enfunc}) is when $\kappa$ and $F$ are large (as compared with $k_BT$)
while their ratio is kept fixed. 

We also emphasize that our model is one where the size of the objects that are wrapped, the cylinders, is large with respect to the microscopic details of the polymer. The effect of 
absorbed objects on a more discrete model has been studied in Ref. \cite{marko}, where in particular the effect of absorbed objects on the effective persistence length was taken into account.

\subsection{Wrapping adhesive energy}

Consider now a cylinder embedded in the plane which is free to move and 
which can become attached to the elastic filament. It is important that  the 
cylinder can move since the position at which the filament is attached to the 
cylinder is not constrained in space. We assume that there is a favorable 
energy of interaction between the cylinder and the filament which is local 
and proportional to the arc-length of the filament touching the cylinder. We further 
assume that in the {\em interaction region} between the filament and the cylinder 
the former follows the surface of the cylinder exactly and that the energy of 
interaction is consequently given by
\begin{equation}
E_i = -\gamma s_c, \label{gamma}
\end{equation}
where $\gamma$ is a line tension parameterizing the affinity of the filament for the cylinder  and $s_c$ is the arc-length of the filament which is attached to the cylinder. If the angle at which the filament 
first touches the cylinder (advancing along the arc-length to the cylinder) 
is $\psi_1$ and the angle at which it leaves is $\psi_2$ (see Fig.  \ref{schematic}, top), 
then the length  of the filament touching the cylinder is $s_c = R|\psi_2-\psi_1|$ and thus
\begin{equation}
E_i=-\gamma R |\psi_2-\psi_1|.
\label{surf1}
\end{equation}
Note that if $\psi_2>\psi_1$ the filament wraps anti-clockwise 
and if $\psi_2<\psi_i$ then it wraps clockwise around the filament. 
As well as having a surface interaction term, there is also a mechanical bending (potential) energy associated with 
the arc-length following the contour of the cylinder. This mechanical energy is  
given by $E_m = E(s_c)$, where the energy functional is defined in Eq. (\ref{enfunc}).
The path that follows the cylinder arriving at angle $\psi_1$ and leaving at angle 
$\psi_2$ is parametrized over $[0,s_c]$ as $\psi_s = 
(\psi_2-\psi_1)s/s_c + \psi_1$. 
Substituting this trajectory into  Eq. (\ref{enfunc}) yields
\begin{equation}
E_m = {\kappa\over 2R}|\psi_2-\psi_1| -FR\ {\rm sgn}(\psi_2-\psi_1)
   \left[\sin({\psi_2})-\sin({\psi_1})\right].
   \label{surf2}
\end{equation}
The total energy of the system is then composed of a  {\em bulk}  elastic term, Eq. (\ref{enfunc}), for the free segments of the filament and a {\em boundary} 
or {\em contact} interaction energy $E_c $ with the cylinder which is a sum of the terms in Eqs. (\ref{surf1}) and (\ref{surf2}), i.e., $E_c=E_i+E_m$. 

One notes that the above energy functional  
is analogous to a 1D Coulomb fluid with a charged boundary, but with an imaginary surface charge \cite{Dean1,Dean2}.

\subsection{Partition function of a filament-cylinder complex}

To determine the conformational equilibrium of a system consisting of an elastic filament with any number
of wrapped cylinders we must evaluate the corresponding partition function. To do this we first define the evolution kernel
$K(\psi,\psi',s)$ which evolves the state of the filament for a distance $s$ measured  along its
arc-length with the boundary condition that $\psi(0) = \psi$ and $\psi(s) = \psi'$. This kernel is
defined by
\begin{equation}
K(\psi,\psi',s) = \int_{\psi(0)=\psi}^{\psi(s)=\psi'} d[\psi] \exp\left(-{\beta}E[\psi] \right)
\label{propag}
\end{equation}
where $\beta=1/k_{\mathrm{B}}T$ and $E[\psi] $ is the elastic energy functional in Eq. (\ref{enfunc}).
It determines the evolution in arc-length $s$ of a  {\em wave function} $f(\psi)$ by
\begin{equation}
\int d\psi' K(\psi,\psi',s)f(\psi') = \exp(-sH)f(\psi),
\end{equation}
where $H$ is the corresponding {\em Hamiltonian operator}
\begin{equation}
H= -{1\over 2\beta \kappa} {d^2 \over d \psi^2} -\beta F\cos(\psi).
\label{Hamilt}
\end{equation} 
The partition function for a {\em single} cylinder, which first becomes attached to  the filament at arc-length
$s=l_1$ measured from the fixed end ${\bf x}(s=0)=0$ (see Fig. \ref{schematic}), is then given by
\begin{widetext}
\begin{equation}
Z = \int d\psi_0 d\psi_1 d\psi_2 d \psi_3 K(\psi_0,\psi_1,l_1)S[\psi_1,\psi_2] 
K(\psi_2,\psi_3, L-l_1 -R|\psi_2-\psi_1|),
\end{equation}
where $\psi(0) = \psi_0$ and $\psi(L) = \psi_3$ 
and the definition of $S$ follows from the contact energy (or the interaction energy between the cylinder and the filament as defined in the preceding Section) 
and is given by
\begin{equation}
S[\psi_1,\psi_2] = \exp\left(\beta|\psi_2-\psi_1|(\gamma R- {\kappa\over 2R}) +
\beta FR\ {\rm sgn}(\psi_2-\psi_1)\left[\sin({\psi_2})-\sin({\psi_1})\right]\right).
\end{equation}
\end{widetext}
The generalization to a system in which the elastic filament interacts with several wrapped cylinders is obvious from the above. 

In the limit of a long elastic filament  where the cylinders are close to its midpoint, 
we make use of the fact that the propagator at large arc-lengths is given in terms of the 
ground-state wave function of the corresponding Schr\" odinger equation
\begin{equation}
K(\psi,\psi', L) = \exp(-E_0 L)\Psi_0(\psi)\Psi_0(\psi'), 
\label{sch_eqn_0}
\end{equation}
where $E_0$ is the ground-state energy of the Hamiltonian $H$, Eq. (\ref{Hamilt}),  and $\Psi_0$ the  corresponding 
wave function, i.e.  $H\Psi_0(\psi) = E_0 \Psi_0(\psi)$. 
The ground-state energy in the ground-state-dominance limit then gives also the partition function in the limit of large arc-lengths.

\subsection{Constrained and free wrapping of cylinders}

Because of specific interactions between the filament and the cylinders the wrapping angles $\psi_1$ and 
$\psi_2$ may not be completely free and constraints may exist that 
reduce their overall degrees of freedom. For instance, in the limit where the adhesion energy
$\gamma$ is very large, the elastic filament  could wrap around the cylinder 
a number of times, {\em via} an escape into the third dimension that does not cost any extra energy, or so we assume in this study.

The simplest constraint is to take the wrapping angle  (see Fig. \ref{schematic}, top) $\psi_2-\psi_1 =\alpha_{1}$ fixed. 
The constrained partition function at a fixed wrapping angle can be computed and then the ensemble where it  varies can be obtained by carrying out the remaining integrations with the 
appropriate statistical weights. This holds for a single wrapped cylinder but can be extended to the case of several cylinders as well.

For a {\em single} cylinder we then define the constrained partition 
function for the ensemble where $\psi_2-\psi_1 = \alpha_1$ is kept fixed as
\begin{eqnarray}
Z(\alpha_1) &=& \int d\psi_0 d\psi_1  d \psi_3 K(\psi_0,\psi_1,l_1)
\exp\big(\beta FR\ {\rm sgn}(\alpha_1)\left[\sin({\psi_1+\alpha_1})-\sin({\psi_1})\right]\big) \times \nonumber\\
&& \qquad\qquad\qquad\qquad\qquad\qquad\quad \times \, K(\psi_1 +\alpha_1,\psi_3, L-l_1 -R|\alpha_1 |).\label{za1}
\end{eqnarray}
The partition function for the ensemble where the contact angle can vary freely is consequently given by
\begin{equation}
Z = \int_{-{{L-l_1}/R}}^{{{L-l_1}/R}} d\alpha_1 Z(\alpha_1) 
\exp\left(\beta |\alpha_1|(\gamma R-{\kappa\over 2R})\right).\label{z1}
\end{equation}
Note that given that the point of first contact between the cylinder and filament is at $s=l_1$, the maximum amount of filament that can be wrapped about the cylinder is $L-l_1$, corresponding to wrapping angles of $\pm (L-l_1)/R$ (clockwise and anti-clockwise). 
When $\alpha_1$ is fixed, the surface terms are periodic in $\psi_1$ and one can 
use standard Mathieu function analysis to carry out the computations. The procedure is described in detail in  Appendix\ref{appendix2}. 

For {\em two} cylinders separated by a distance $l$, where the first one is located an arc-length $l_1$ away from the 
fixed end and with contact angles $\alpha_1$  and $\alpha_2$ (see Fig. \ref{schematic}, bottom),  we have in complete 
analogy with above 
\begin{eqnarray}
Z(\alpha_1,\alpha_2)&=&\int d\psi_0 d\psi_1  d \psi_3 d\psi_5\  K(\psi_0,\psi_1,l_1)
{\cal S}_{R_1}(\psi_1, \alpha_1) K(\psi_1 +\alpha_1,\psi_3, l)\times \nonumber\\
 &&\qquad\qquad\qquad\qquad\qquad\qquad\qquad\quad\times\, {\cal S}_{R_2}(\psi_3, \alpha_2) K(\psi_3+\alpha_2,\psi_5, L_{12}),
\end{eqnarray}
where
\begin{equation}
L_{12} = L-l-l_1-R_1|\alpha_1|-R_2|\alpha_2|
\end{equation}
is the arc-length between the right, free, end of the chain and the rightmost cylinder (see Fig. \ref{schematic}, bottom), and
\begin{equation}
{\cal S}_R(\psi, \alpha) = \exp\left( \beta FR\ {\rm sgn}(\alpha)\left[\sin({\psi+\alpha})-\sin({\psi})\right] \right),
\end{equation}
where $R_1$ is the radius of cylinder 1 and $R_2$ is that of cylinder 2.

\subsection{Constrained arc-length separation between wrapped cylinders}

\begin{figure}[t]
   \centering
  \includegraphics[width=9cm]{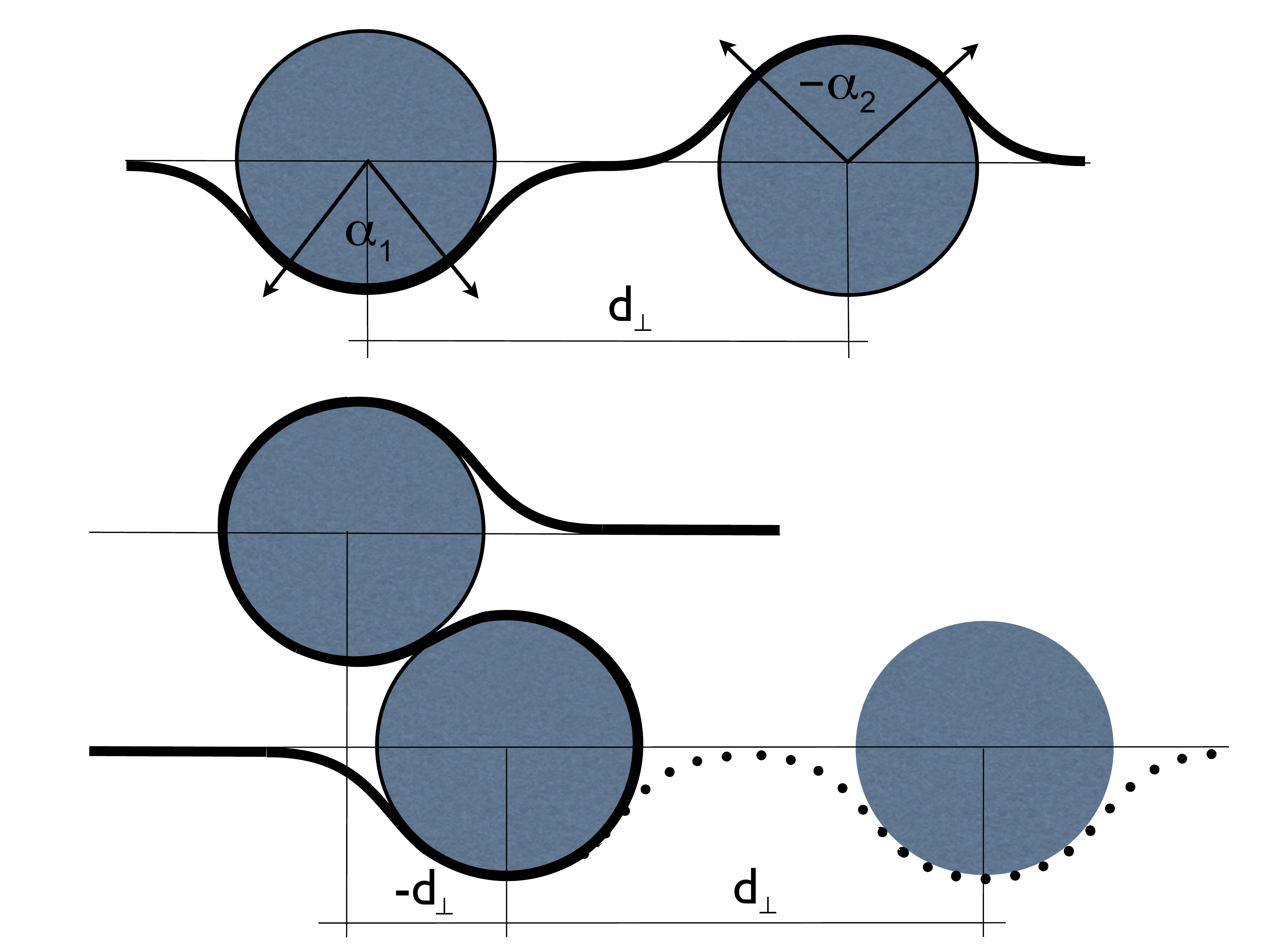}
  \caption{Top: Schematic view of the wrapping of two cylinders in an antisymmetric mode where the wrapping angles for 
the two cylinders differ in sign. Bottom: the looped configuration (full line) of two wrapped cylinders corresponds 
to a negative value of the horizontal projected separation $d_\bot$ and the extended configuration (dotted line) 
with positive $d_\bot$. }
  \label{schematic2}
\end{figure}

Now consider the limit of a long filament where the cylinders are close to the midpoint, i.e., we take $l_1$ and 
$L-l_1$ large while keeping  the arc-length between the final point of the filament touching cylinder 1 and the first touching cylinder 2, $l$, constant (see Fig. \ref{schematic}, bottom). These are all lengths along the chain (arc-length) which are fixed, if we wish to keep the physical distance in a given direction fixed we must consider another ensemble; we shall discuss this later. In this limit (corresponding to the limit of 
ground state dominance, Eq. (\ref{sch_eqn_0})),  we find, up to an overall  factor,
\begin{equation}
Z(\alpha_1,\alpha_2)=\exp\left(-E_0 L_{12}\right) \int d\psi_1  d \psi_3   
 \Psi_0(\psi_1){\cal S}_{R_1}(\psi_1, \alpha_1)  K(\psi_1 +\alpha_1,\psi_3, l)  
  {\cal S}_{R_2}(\psi_3, \alpha_2)  \Psi_0(\psi_3+\alpha_2).
\end{equation}
An interesting case emerges when the two wrapping angles are equal and correspond to complete single wrapping,  
$\alpha_1=\alpha_2=2\pi$. The $l$ dependent part of the partition function is then given by
\begin{equation}
Z(l) = \exp(E_0l)\int d\psi_1 d\psi_3 \Psi_0(\psi_1) K(\psi_1 +2\pi,\psi_3, l) \Psi_0(\psi_3+2\pi),
\end{equation}
and using the periodicity this gives
\begin{equation}
Z(l) = \exp(E_0l)\int d\psi_1 d\psi_3 \Psi_0(\psi_1) K(\psi_1,\psi_3, l) \Psi_0(\psi_3).
\end{equation}
The same result is found if one assumes that the wrapping is antisymmetrical, $\alpha_2=-\alpha_1=-2\pi$. 
The symmetry of wrapping between the two cylinders will be addressed furthermore below. The difference between 
symmetric and antisymmetric wrapping is obvious from Figs. \ref{schematic} and \ref{schematic2}.

\subsection{\label{wrapping_th} Wrapping-unwrapping transition of cylinders on an elastic filament } 

We first consider the case of a single adsorbing cylinder. The elastic filament is of length $L$ and a wrapped cylinder is positioned at $l_1$. We restrict ourselves to the case where both $L$ and $l_1\to\infty$, assuming that $l_1$ is not fixed so that the cylinder can attach anywhere and consider the case where the elastic filament  can wrap on the cylinder any number of times.

In this case, the expression in Eq. (\ref{za1}) becomes, up to a constant prefactor, independent of $\alpha$:
\begin{equation}
Z(\alpha) = \exp\left(-E_0(L-R|\alpha|)\right)f(\alpha), \qquad {\rm with} \qquad f(\alpha) = \int_0^{2\pi}d\psi \, \Psi_0^2(\psi) {\cal S}_{R}(\psi, \alpha),
\label{inetgen}
\end{equation}
and the function $f(\alpha)$ is clearly bounded. From Eq. (\ref{z1}) the partition function for the ensemble with variable $\alpha$ 
is obtained by integrating Eq. (\ref{inetgen}) over $\alpha$ from $-{{L}/ R}$ to $+{{L}/ R}$, i.e.
\begin{equation}
Z = \exp(-LE_0) \int_{-{{L}/ R}}^{{{L}/ R}} d\alpha ~\exp\left(
\beta |\alpha|(\gamma R-{\kappa\over 2R}+{R E_0\over \beta})\right) f(\alpha).
\end{equation}
Defining 
\begin{equation}
\Delta E = \gamma R-{\kappa\over 2R}+{R E_0\over \beta},
\label{ptran}
\end{equation}
we see that the elastic filament will then wrap around the cylinder a macroscopic number of times (in the sense that the total length wrapped around the filament will be of the order of the filament length) if $\Delta E > 0$. However, for $\Delta E < 0$ the filament will have only a microscopic length wrapped around the cylinder. The equation $\Delta E = 0$
therefore defines a {\em wrapping transition} in the phase diagram of variables $\beta$ and $F$  as a consequence of  competition between the wrapping energy of the cylinder and the configurational entropy of the chain. 

In the zero-temperature limit where $\beta\to\infty$ the ground state of the filament configuration is a 
straight line in the direction of the applied force and so without a 
cylinder
\begin{equation}
Z(L) \approx \exp(-LE_0)=\exp(\beta FL),
\end{equation}
or $E_0(T=0) = -\beta F$. Therefore, if $\gamma R-\kappa/ 2R-F R >0$,
the system is in the wrapped state, otherwise the filament unwraps from the cylinder. This 
conclusion is easy to deduce from purely energetic arguments for large values of 
the wrapping angle $\alpha$. 




We next consider several cylinders in the plane and assume that the filament 
can wrap around all of them without impediment. The wrapping will induce effective 
interactions between the cylinders that can be either repulsive, 
attractive or non-monotonic. Wrapping transitions with effective interactions between cylinders can be viewed as a 
model for nucleosomal wrapping \cite{Arneodo1,Arneodo2,hoda4}. However, the details of the most general case remain  to be elaborated. 

For clarity we briefly describe the system with three cylinders: cylinder $1$ at position $l_{01}$ from the left end of the chain, cylinder $2$ separated from $1$ by a distance $l_{12}$, cylinder $3$ separated from $2$ by a distance $l_{23}$ and with contact angles $\alpha_1$, $\alpha_2$ and $\alpha_3$. The partition function can be written as
{\bacol
\begin{eqnarray}
&&Z(\alpha_1,\alpha_2, \alpha_3) = \int d\psi_0 d\psi_1  d \psi_3 d\psi_5d\psi_7 
 K(\psi_0,\psi_1,l_{01}){\cal S}(\psi_1, \alpha_1)
  K(\psi_1 +\alpha_1,\psi_3, l_{12})  {\cal S}(\psi_3, \alpha_2)\times \nn\\
&& \times K(\psi_3 +\alpha_2,\psi_5, l_{23})  {\cal S}(\psi_5, \alpha_3)
   K(\psi_5+\alpha_3,\psi_7,L- (l_{01} +l_{12}+ l_{23} +R_1|\alpha_1|+R_2|\alpha_2|+R_3|\alpha_3|)),\nonumber\\
\end{eqnarray}
}
with the general definition 
\begin{eqnarray}
{\cal S}(\psi_{2i+1}, \alpha_i) &=& \exp\left(\beta FR_i\ {\rm sgn}(\alpha_i)\left[
\sin({\psi_{2i+1}+\alpha_i})-\sin({\psi_{2i+1}}) \right]\right),
\end{eqnarray}
where $i$ (for $i=0,1,2$) is the index of the wrapped cylinder. 

The general system for $N$ particles bears some overall similarity to the 1D Tonks gas, although 
the effective interactions between the cylinders are more complicated. In order 
to make this system applicable to the problem of nucleosome wrapping around DNA, 
it may be necessary to insert a chemical potential term for the cylinders 
similar to the case of a grand-canonical Tonks gas \cite{Arneodo1,Arneodo2}. The 
main difference between the two models is, however, in the fact that the 
interaction between the wrapped cylinders along the elastic filament  can be much more 
complicated than in the Tonks case and could in principle lead to  ``phase 
separation" without imposing any excluding energy barriers along the chain \cite{hoda4}.

\section{\label{one_wa} Unwrapping transition: One cylinder}

We now apply the general theory derived in the previous Sections to the problem 
of the unwrapping transition for a single wrapped cylinder.

To accommodate the wide range of values that the variables take in the physical 
systems to which our theory would be applicable, it is convenient to recast the 
calculation in terms of equivalent dimensionless variables. We consider an 
elastic filament  of persistence length $L_p$, fixed at one end and pulled from 
the other end with a force $F$. The wrapped cylinders are all of radius $R$. We 
define the following dimensionless variables
\begin{equation}
\mu = \frac{2L_p}{R},\; f=\beta R F,\; \sigma=\beta\gamma R,\; \varepsilon = ER, 
\label{dimless}
\end{equation}
where $L_p=\beta \kappa$ and $\gamma$, the wrapping energy of a cylinder per unit of arc-length, and other parameters have been defined before, Eqs. (\ref{eq:bending-energy})-(\ref{gamma}). In terms of these variables the Hamiltonian can be written as $H= H'/(\mu R)$
and thus $H'$ has eigenvalues which can be written in the form of the Schr\" dinger  equation 
\begin{equation}
H'\psi_m = \mu \epsilon_m \psi_m \label{sch_eqn}
\end{equation}
where $\epsilon_m = E_m R$. 

As an illuminating example let us consider a specific case with the dimensionless quantities 
$\mu=1$ and $\sigma = 0.5$. Assuming cylinder radius of $R=2$~nm, this gives actual parameter
values as $L_p = 1$~nm (persistence length), $\kappa = k_BT \times L_p \simeq 4\times 10^{-30}\mbox{J.m}$ 
(bending stiffness) at $T=300$~K, 
and $\gamma=1$~pN (line tension for cylinder-filament adhesion). 

For the single cylinder wrapping transition, a rough estimate for its occurrence 
can be obtained in the following way. In Section \ref{wrapping_th}, it was argued 
that for large $\beta$ (low temperature) $ E_0 \sim \beta F$ and that the 
wrapped phases  occurs for forces which obey $\Delta E > 0$, Eq. \ref{ptran}. 
For scaled quantities as introduced above
this inequality becomes
\begin{equation}
f~<~\sigma-\frac{\mu}{4} \qquad {\rm and~thus} \qquad \sigma'~\equiv~\sigma - 
\frac{\mu}{4}~>~0\;.  \label{f1}
\end{equation}
Alternatively, we can then consider the Schr\" dinger  equation, Eq. \ref{sch_eqn} (see also Eq. (\ref{Hamilt})), as an approximate oscillator 
equation and get an improved estimate 
$ E_0~=~-\beta F + \sqrt{{F}/{(4\kappa)}}$. Using Eq. (\ref{f1}) above, the condition for the wrapping  transition then becomes
\begin{equation} 
f - \sqrt{{f}/{2\mu}}-\sigma' < 0\;,
\label{f2}
\end{equation}
This can be recast as a quadratic equation in $\sqrt{f}$ and
the condition can be satisfied if the discriminant is positive, i.e.
\begin{equation}
\sigma'~>~-\frac{1}{8\mu}\;\qquad {\rm and~thus} \qquad\sigma~>~\frac{\mu}{4}-\frac{1}{8\mu}.
\label{wcond}
\end{equation}
In the limit where $\mu$ is large, the above expression reduces to the condition in 
Eq. (\ref{f1}). There will of course be corrections to this form due to the 
non-harmonic terms in the effective potential in Eq. (\ref{sch_eqn}).  
Nevertheless, from both arguments it follows that a sufficient condition for the 
transition to occur is $\sigma > \mu/4$. If we return to original variables we see that the second
condition of Eq. ({\ref{wcond}) gives
\begin{equation} 
\gamma > \gamma',\qquad {\rm where}\quad \gamma' = {1\over 2} {\kappa\over R^2} -{(k_B T)^2\over 16 \kappa}.
\label{Twcond}
\end{equation}
In the limit of zero temperature this result agrees with that of Weikl \cite{weik} for a cylinder and membrane. The 
interpretation of the zero-temperature result is simply that the adhesive energy must overcome the bending energy 
to make the bound state stable. We see that the effect of temperature is to diminish the minimum adhesion energy 
necessary to obtain a bound phase. At first sight this appears counter-intuitive and so we investigate the
effect of temperature below in more detail.  

We first present an exact calculation of the wrapping transition in 
a particular case. The mean value of $\alpha$, i.e. $\la \alpha \ra$, is calculated for fixed $(\mu,\sigma)$ 
as a function of the reduced force $f$. Whilst taking the filament to be of infinite extent we restrict the 
range of $\alpha$ to be $-\alpha_{max} < \alpha < \alpha_{max}$ and take $\alpha_{max}=100$. For small 
$f$ we  expect $\la \alpha \ra$ to be large, which corresponds to the elastic filament maximally wrapped 
onto the cylinder: $|\alpha|=\alpha_{max}$. As $f$ is increased the system will go through a transition from the 
wrapped to the unwrapped configuration.

To solve the corresponding Schr\" dinger  equation, the range of $x$ is chosen to be an 
integer multiple $N$ of $2\pi$, where $N$ would then be the number of lattice sites in the 
Bravais lattice for the discretized Hamiltonian. The Schr\"odinger equation  (Eq. \ref{sch_eqn}) 
is then recast as a matrix equation with imposed periodic boundary conditions. 
This will lead to a band structure with the Brillouin zone and the number of 
Bloch states per band determined by $N$.  Since the solution of this problem requires {\em only} strictly 
periodic eigenfunctions we choose $N=1$ and the range of $x$ as $2\pi$. The 
eigenvalues are then equal to $\mu\epsilon_m$.

\begin{figure}[t!]
   \centering
  \includegraphics[width=9cm]{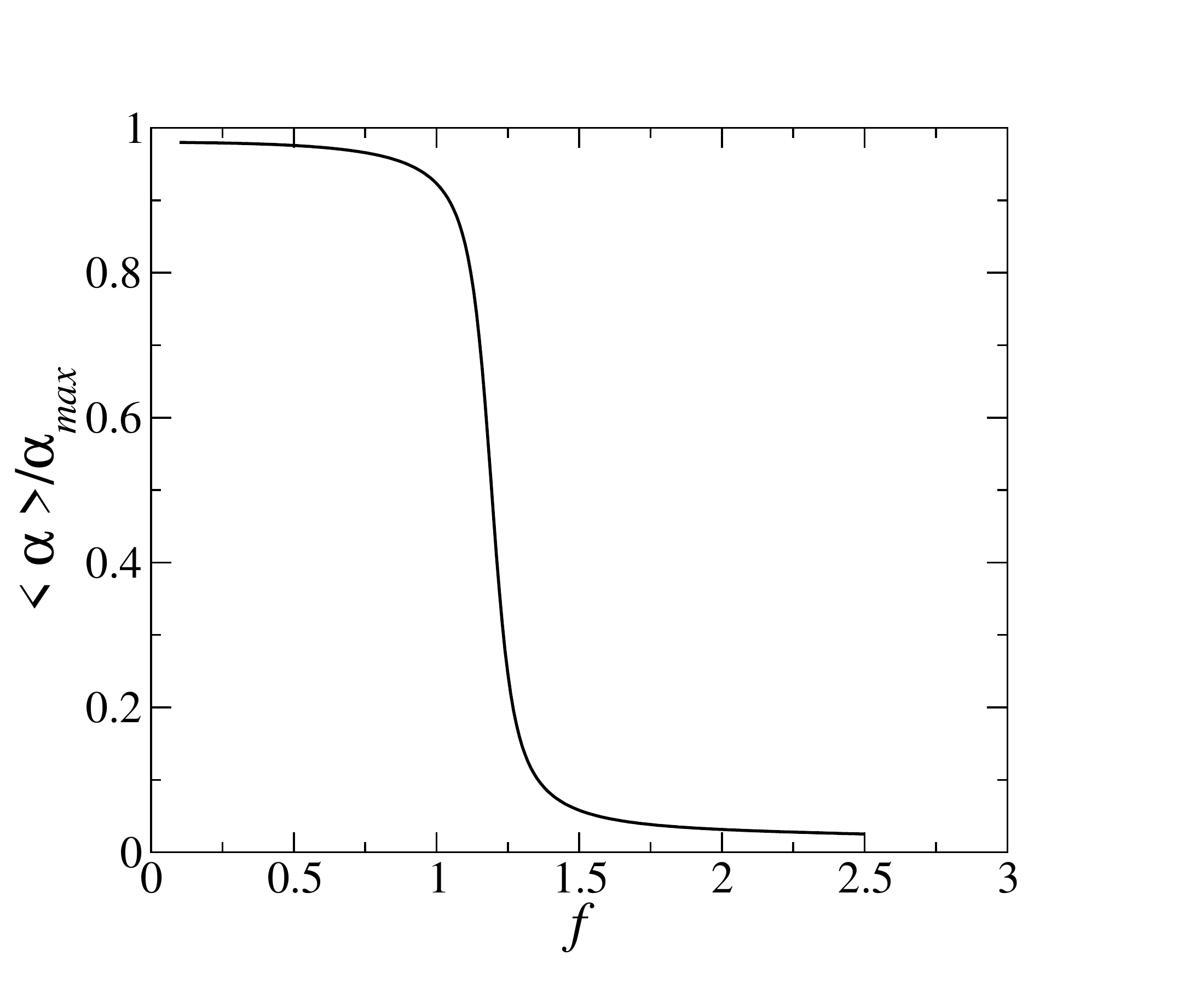}
  \caption{The wrapping transition of a single cylinder on an elastic filament. Exact solution for the 
    average wrapping angle ratio $\mathopen< \alpha \mathclose>/\alpha_{max}$ is shown as a function of the  
     dimensionless external tension $f=\beta R F$ for fixed $\mu=1$ and $\sigma=0.75$.}
  \label{winding}
\end{figure}

In Fig. \ref{winding} we show the results for the mean wrapping angle $\mathopen< \alpha \mathclose>$, scaled by $\alpha_{max}$,  as a function of the external dimensionless force $f$ for fixed  $\mu=1$ and $\sigma=0.75$. The wrapping transition, corresponding to the maximum  in the derivative ${\partial \la \alpha \ra}/{\partial f}$, then occurs for the critical reduced force  $f_c \simeq 1.2$ which, for the parameters stated above (i.e., using $L_p = 1\mbox{~nm},~R = 2\mbox{~nm}$) gives $F_c \simeq 2.4\mbox{~pN}$. Compare this with the corresponding 
estimates from
\begin{itemize}
\item[(i)] Eq. (\ref{f1}). We find 
\[
f_c~=~\sigma-\mu/4~=~0.5~~\longrightarrow~~F_c = 1\mbox{~pN}.
\]
\item[(ii)] Eq. (\ref{f2}). We solve the quadratic equation (for $\mu=1$ and $\sigma=0.75$) to get 
\[
f_c~=~1.31~~ \longrightarrow~~F_c = 2.67\mbox{~pN},
\]
\end{itemize}
Obviously, in order to get an accurate prediction for the critical force it is necessary to solve the full 
Schr\" dinger  equation as we did above.  Nevertheless, the harmonic approximation in Eq. (\ref{f2}) is in acceptable agreement 
with this exact result.

\begin{figure*}[t!]\begin{center}
        \begin{minipage}[b]{0.485\textwidth}
        \begin{center}
                \includegraphics[width=\textwidth]{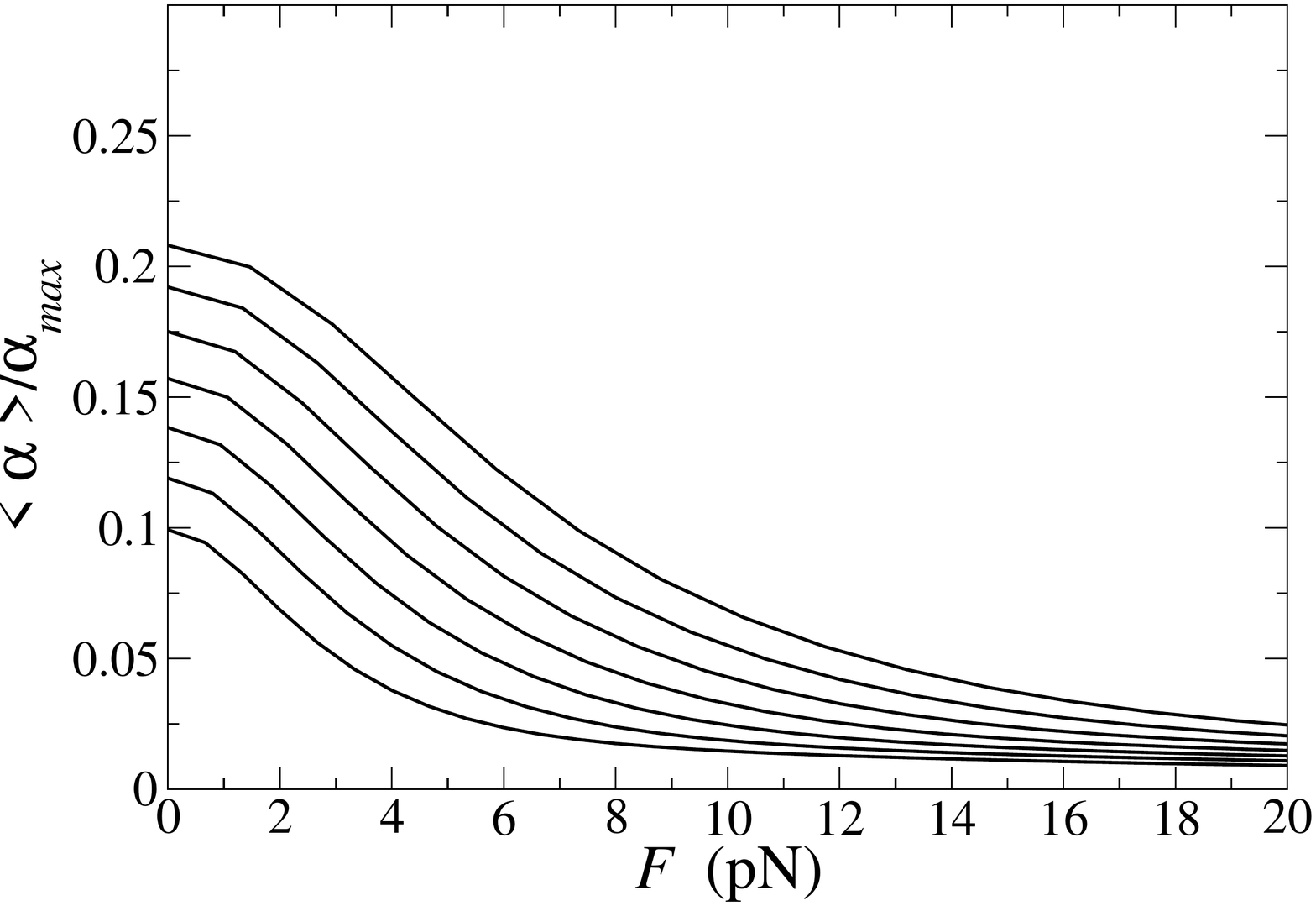}
        \end{center}\end{minipage} \hskip0.25cm
        \begin{minipage}[b]{0.49\textwidth}\begin{center}
                \includegraphics[width=\textwidth]{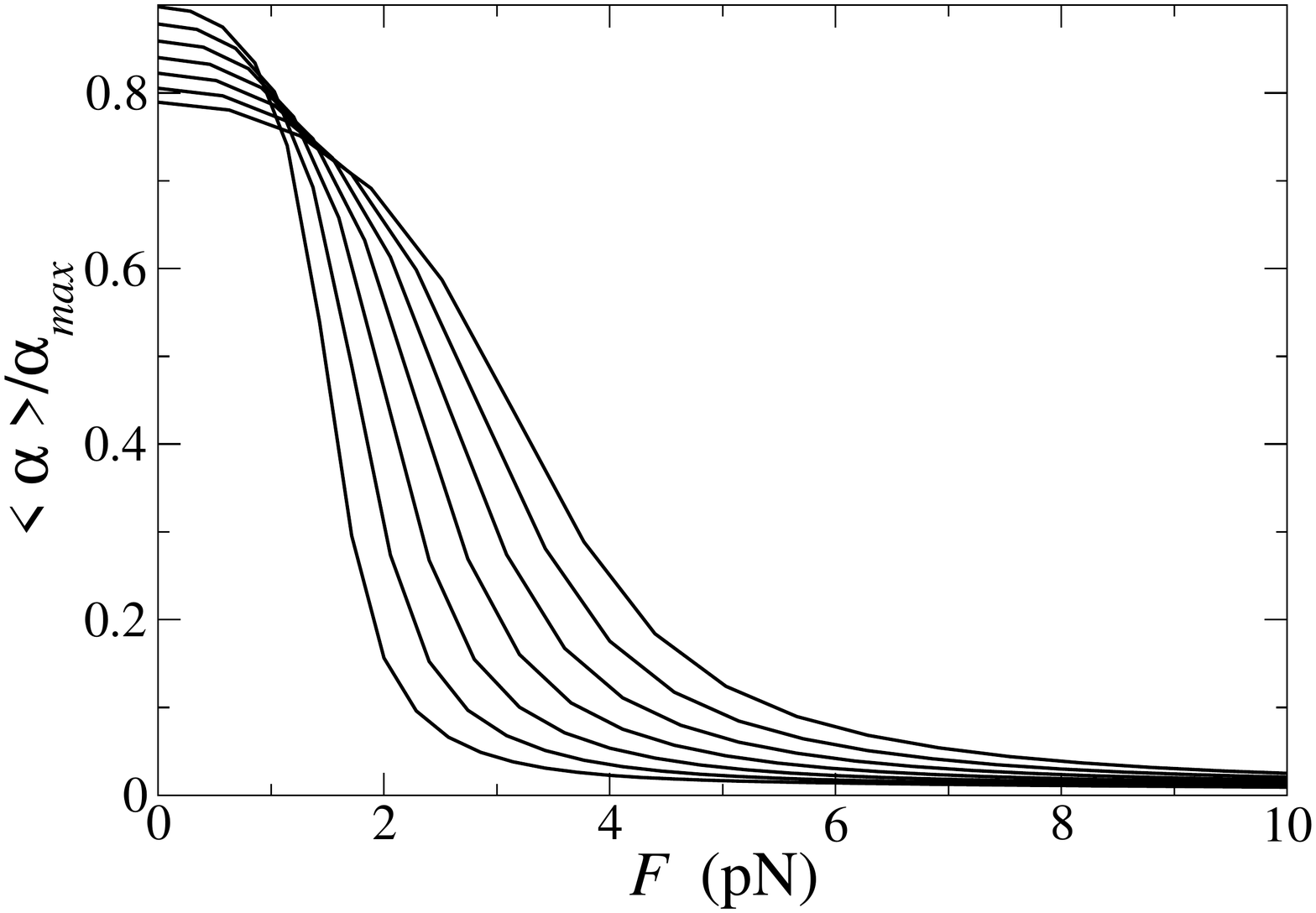}
        \end{center}\end{minipage} \hskip0.25cm
\caption{The wrapping transition as a function of the actual force $F$ and for temperatures $T = 210$~K up to 390~K  in steps of 30~K. The physical values for the parameters are given in the text. In both cases the curves move right as $T$ increases. Left pane: we have $\sigma/\mu = 0.15$, $\sigma' < 0$. The wrapping transition is relatively  smooth and as $T$ increases it becomes stronger and moves to larger values of $F$. Right pane:  $\sigma/\mu = 0.35$, $\sigma' > 0$.  The wrapping transition is much stronger and sharper as compared with the former case and,  as predicted, the transition moves to larger values of $F$ as $T$ increases. It is this behavior in both cases  which might be considered counter-intuitive.}
\label{Winding_2D}
\end{center}\end{figure*}

In order to highlight the effect of temperature it is important to consider the 
dependence of $\langle \alpha \rangle$ on the force $F$ rather than on the dimensionless 
reduced force $f$ because the latter contains a hidden dependence on $T$ which would obscure 
the effect we are studying. In the harmonic approximation, we have that the critical force $F_c$ is 
predicted to be 
\begin{equation}
\sqrt{F_c} = \frac{k_BT}{4\sqrt{\kappa}} + \left(\gamma-\gamma'\right)\;. \label{Fc_harmonic}
\end{equation}
We investigate two cases and take typical values for the parameters similar to those stated at the 
beginning of this Section, i.e., $R = 2$~nm, $L_p = 1$ nm, $\kappa = k_BT \times L_p$ and the temperature is taken as $T = 300$~K, so that $\mu = 1$. We study two cases 
corresponding to $\sigma/\mu =0.15$ and $0.35$, respectively. We also note that these choices respectively give $\sigma' > 0$ and $\sigma' < 0$, which 
allows us to test the significance of the inequality in Eq. (\ref{wcond}). We study a wide range of 
temperatures above and below $T=300$~K which is necessary to reveal the $T$-dependence of 
the wrapping transition. Of course, such a wide temperature range does not occur {\em in vivo} but 
our aim is to test the prediction inferred from Eq. (\ref{Fc_harmonic}), namely that the critical force for
the wrapping transition increases, rather than decreases, with increasing $T$. The values for $T$ considered
are between $T= 210$ and $390$~K in intervals of $30$~K, and the results are shown in Fig. \ref{Winding_2D} for the
above two cases:

\begin{itemize}
\item[(i)] $\sigma/\mu = 0.15$, $\sigma' < 0$: This corresponds to relatively weak adhesion coefficient 
$\gamma = 0.30\mbox{~pN}$
and whilst we expect a wrapping transition to occur, we expect it to be relatively smooth and, 
according to Eqs. (\ref{Twcond}) and (\ref{Fc_harmonic}), to turn on more strongly as $T$ increases and for 
the critical force to increase as $T$ increases. We see these features in Fig. \ref{Winding_2D}, left pane. We 
conclude also that even though $\sigma' < 0$ the wrapping transition does occur. For $T=300$~K the harmonic 
approximation in Eq. (\ref{Fc_harmonic}) predicts $F_c = 0.52\mbox{~pN}$. The transition is not sharp and so 
it is not feasible to read a specific value for $F_c$ from the graph but it is clear that this prediction 
is very much on the low side. 

\item[(ii)] $\sigma/\mu = 0.35$, $\sigma' > 0$: The adhesion coefficient 
$\gamma = 0.70\mbox{~pN}$ is larger and, 
in this case, we expect the transition to be stronger and sharper than in case (i). This is seen in Fig.
\ref{Winding_2D}, right pane. As $T$ increases the transition moves to larger critical force as predicted.  
For $T=300$~K the harmonic approximation in Eq. (\ref{Fc_harmonic}) predicts $F_c = 1.37\mbox{~pN}$. From the graph
the value of $F_c$ follows as $F_c \sim 3\mbox{~pN}$, twice as big as the harmonic prediction.
\end{itemize}
We have thus verified that the effect of increasing temperature $T$ is to increase the critical force at which
the wrapping transition occurs and  this feature is clearly seen in Fig. \ref{Winding_2D}(ii). In other words we can conclude that increasing the temperature causes the wrapping transition to occur at larger force and the corollary is that for fixed force increasing the temperature can actually cause the transition: a somewhat surprising result! 

These effects can be understood by noting that it is the ground-state energy $E_0$ for the full Schr\" dinger  equation
which contributes in Eq. (\ref{inetgen}) and this is clearly $T$-dependent and different from $E_0$ for the 
harmonic approximation. As $T$ increases the potential becomes narrower leading to $E_0$ increasing with $T$. 
This leads to a smaller threshold, $\gamma'$, as defined in Eq. (\ref{Twcond}).

\section{\label{unbinding} Unwrapping transition: two cylinders}

We now study the unwrapping transition in a system of two cylinders of radius $R$, labelled by $i=1,2$, wrapped by a length of filament. There are two cases which we consider.

The first case (case I) consists of the two cylinders wrapped by a fixed length of filament. The relevant variables, shown in Fig. \ref{schematic}, are as follows: 
\begin{itemize}
\item[-] The total angle, $\alpha_i$, wrapped by filament for the $i$-th cylinder. These are both fixed (quenched) for $i=1, 2$.
\item[-] The length, $l$, of filament directly between the two cylinders, i.e., from the exit of the first cylinder to the entrance of the second, which is also fixed.
\end{itemize}
We encode the configuration of the cylinders by measuring the expectation value of $d_\perp$, the projected horizontal displacement between the centers of the two cylinders. The maximum projected horizontal displacement is given by $d_{max}=l+2R$. We shall deal with this case in Section \ref{two_constrained} below. 

The second case (case II)  consists of the two cylinders pinned a distance $l^\prime$ apart 
along a filament, which are wrapped by a dynamical (or annealed) length of filament. The relevant variables are:
\begin{itemize}
\item[-] The length $l^\prime$ of filament between the sites of pinning of the two cylinders. This is fixed (quenched).
\item[-] The total angle, $\alpha_i$, wrapped by filament for the $i$-th cylinder. These are dynamical (annealed) and free to change.
\item[-] The angle, $\alpha_i^\prime$, wrapped by the {\em internal} length of filament (initially of length $l^\prime$) between cylinders for the $i$-th cylinder. These are also dynamical; they take into account the ways in which the filament can wind. 
\item[-] The length, $l$, of filament directly between the two cylinders, i.e.,  from the exit of the first cylinder to the entrance of the second. It is now dynamical being determined by 
$l=l^\prime-R(\alpha_1^\prime+\alpha_2^\prime)$.
\end{itemize}

Again we encode the configuration of the cylinders by measuring the expectation value of $d_\perp$, the projected horizontal displacement between the centers of the two cylinders. The maximum projected horizontal displacement is given now by $d_{max}=l^\prime+2R$. We shall deal with this case in Section \ref{two_unconstrained} below. 

\subsection{\label{two_constrained} Two cylinders: constrained wrapping}

In the first case (case I) where the wrapping angles are fixed, we write the 
constrained partition function of this system as
\begin{widetext}
\begin{equation}
Z(\alpha_1,\alpha_2,l)~= \int d\psi_1d\psi_2\, \langle 0|\hat{O}(\psi_1,\psi_1+\alpha_1) 
\exp\left(-Hl/R\mu\right)\hat{O}(\psi_2,\psi_2+\alpha_2)|0\rangle,
\label{conpart}
\end{equation}
\end{widetext} 
where we assume that the ground state dominates the external regions of the elastic filament, i.e., regions outside the part bounded by the two wrapped cylinders.  The 
operator insertion $\hat{O}$ corresponds to the wrapping of the elastic filament  around a 
cylinder and is given by
\begin{equation}
\hat{O}(\psi,\psi+\alpha) = C_{\alpha}(\psi)|\psi\ra\la\psi+\alpha|, 
\end{equation}
where
\begin{equation}
C_{\alpha}(\psi) = \exp\left( {|\alpha|\left[\sigma+\epsilon_0\right]+
                  f\;{\rm sgn}(\alpha)\left[\sin{(\psi+\alpha)}-\sin{\psi}\right]}\right).
\label{Calpha}
\end{equation}
This gives for the constrained partition function Eq. (\ref{conpart})
\begin{widetext}
\begin{equation}
Z(\alpha_1,\alpha_2,l) = 
\int d\psi_1d\psi_2 \Psi_0(\psi_2)C_{\alpha_1}(\psi_1)\sum_m e^{{-\epsilon_m l/R}}
 P_m(\psi_1+\alpha_1,\psi_2) C_{\alpha_2}(\psi_2)\Psi_0(\psi_2+\alpha_2),
\end{equation}
\end{widetext}
where
\begin{eqnarray}
P_m(\psi_1,\psi_2)=\Psi_m(\psi_1)\Psi_m(\psi_2).
\end{eqnarray}
The functions $P_m$ and the exponential in front of the $P_m$ can be pre-computed 
outside any nested integration loops for fixed $\alpha_i$, leading to a quick and straightforward  
evaluation of the above expression. We are interested in the average horizontal distance between 
the centers of the two cylinders which is given by
\begin{widetext}
 \begin{eqnarray}
\la d_\bot \ra&=&\frac{1}{Z(\alpha_1,\alpha_2,l)}\int\! ds \int\!  d(\sin{\psi_s})\! \! 
 \int \! d\psi_1d\psi_2 \bigg( \langle 0|
 \hat{O}(\psi_1,\psi_1+\alpha_1)e^{{-Hs/R\mu}} |\psi_s\rangle\times\nonumber\\
 &&\qquad \times\, \langle 
 \psi_s| e^{{-H(l-s)/R\mu}}\hat{O}(\psi_2,\psi_2+\alpha_2)|0\rangle + R\,{\rm sgn}(\alpha_1)\la\sin(\psi_2)\ra-R\,{\rm sgn}(\alpha_2)\la\sin(\psi_3)\ra\bigg).  \nonumber\\
 \label{dublerw}
 \end{eqnarray}
 \end{widetext}
A form suitable for numerical computations is reworked in Appendix \ref{appendix1}.

We remark at this point that in our model nothing stops the cylinders from 
passing through each other. Therefore the inter-cylinder force for small 
horizontal separations has to be interpreted only up to the limit of cylinders 
actually touching. The definition of this point depends on the symmetry of the 
wrapping, i.e., it differs when $\alpha_1 = \alpha_2$ and when $\alpha_1 = 
-\alpha_2$. These are the two cases that we have designated symmetric and antisymmetric, respectively.

The wrapping mediated interactions derived in this way should then 
be added to the hard core 1D Tonks gas model \cite{comm}, which is what was only taken into 
account in the studies of the positional distribution of nucleosomes along the 
genome \cite{Arneodo1,Arneodo2}. We do not delve into this problem specifically 
here, but plan to address it elsewhere.

\begin{figure*}[t!]\begin{center}
	\begin{minipage}[b]{0.485\textwidth}
	\begin{center}
		\includegraphics[width=\textwidth]{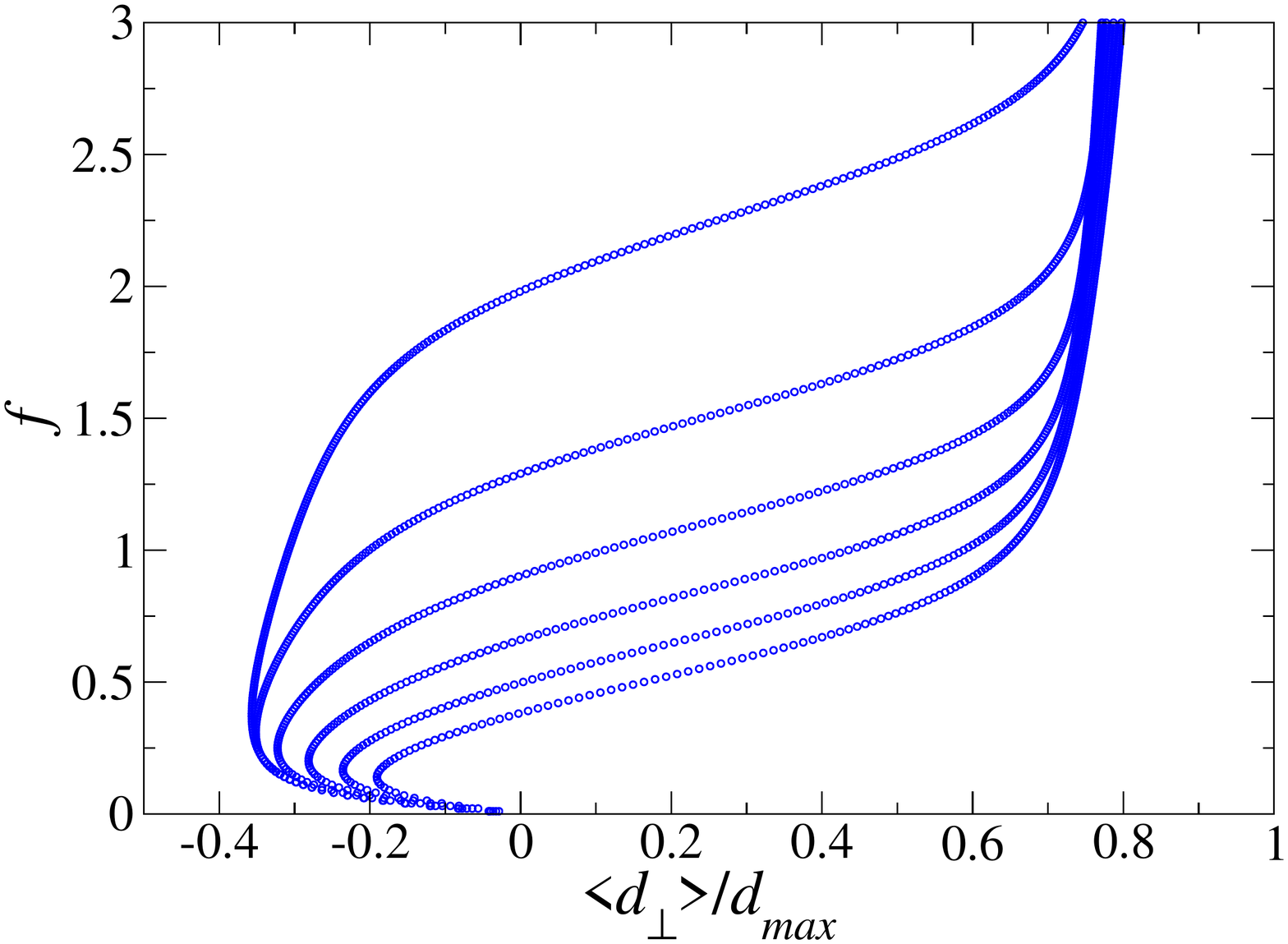}
	\end{center}\end{minipage} \hskip0.25cm
	\begin{minipage}[b]{0.49\textwidth}\begin{center}
		\includegraphics[width=\textwidth]{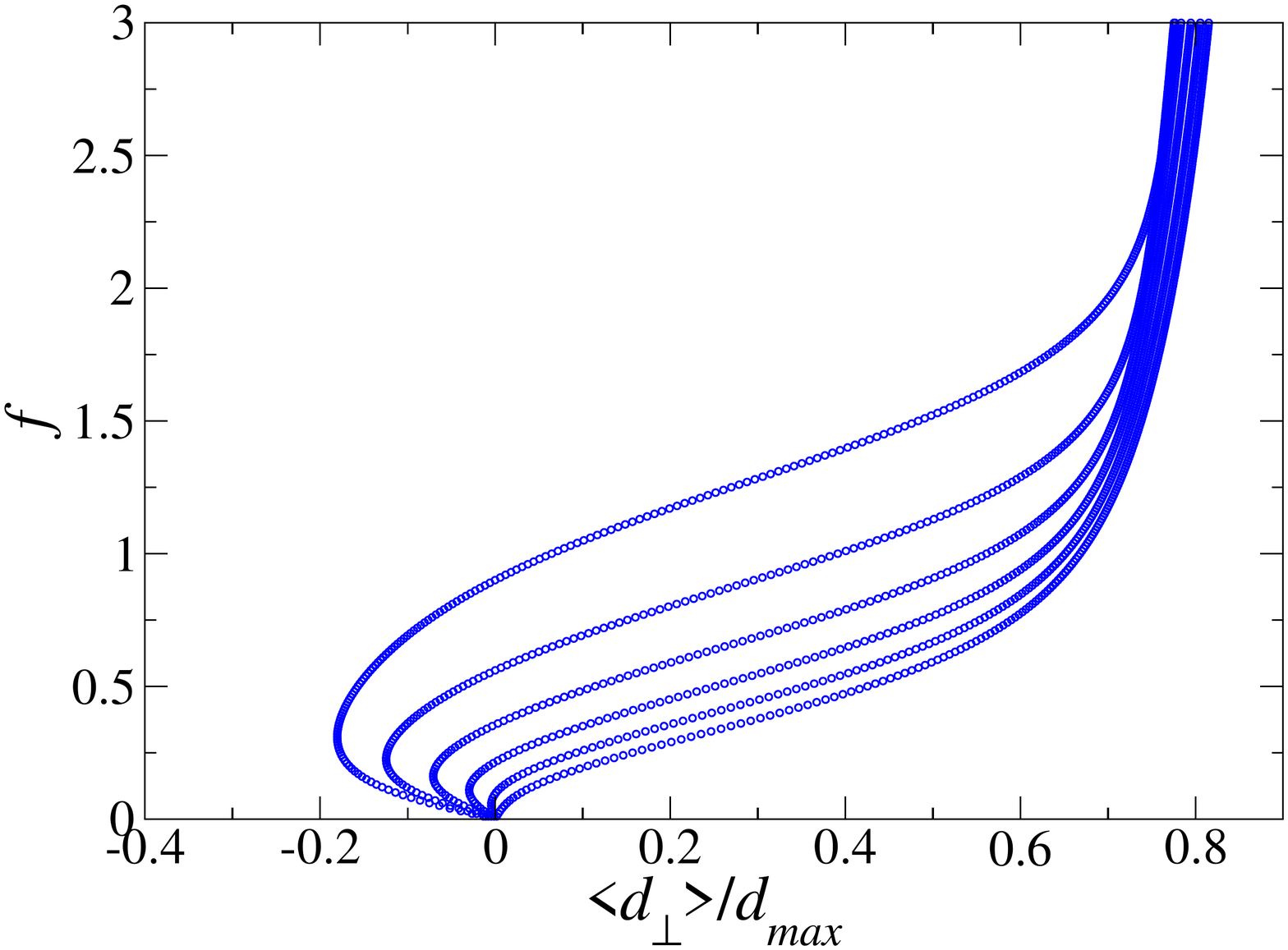}
	\end{center}\end{minipage} \hskip0.25cm	
\caption{Dependence of the average horizontal separation $\la d_\bot \ra$ (normalized to $d_{max}=l+2R$) on 
the external dimensionless tension $f=\beta R F$ for different sets of 
parameters in the case of constrained symmetric wrapping with $\alpha_1=\alpha_2=\pi$. 
Left pane:  $\mu=10$ (high rigidity). Right pane: $\mu=5$  (low rigidity). Negative values of the average horizontal 
separation indicate the presence of a looped phase, i.e., the cylinder at 
larger distance along the elastic filament  lies to the left of the other cylinder. Curves from left 
to right correspond to $l=nR$ for integer $n=3,\ldots,8$.}
\label{fig1}
\end{center}\end{figure*}

\begin{figure*}[t!]\begin{center}
	\begin{minipage}[b]{0.485\textwidth}
	\begin{center}
		\includegraphics[width=\textwidth]{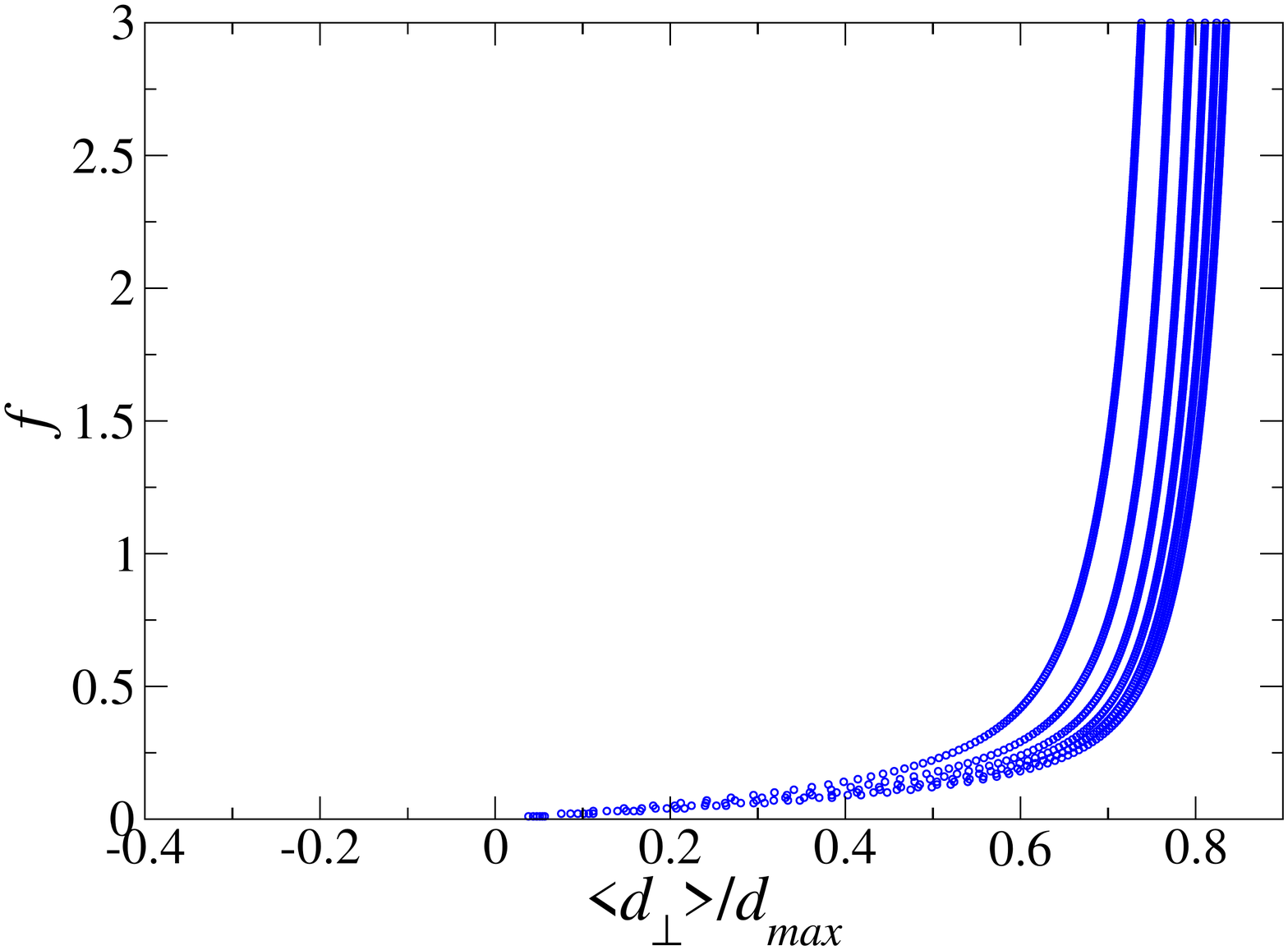}
	\end{center}\end{minipage} \hskip0.25cm
	\begin{minipage}[b]{0.49\textwidth}\begin{center}
		\includegraphics[width=\textwidth]{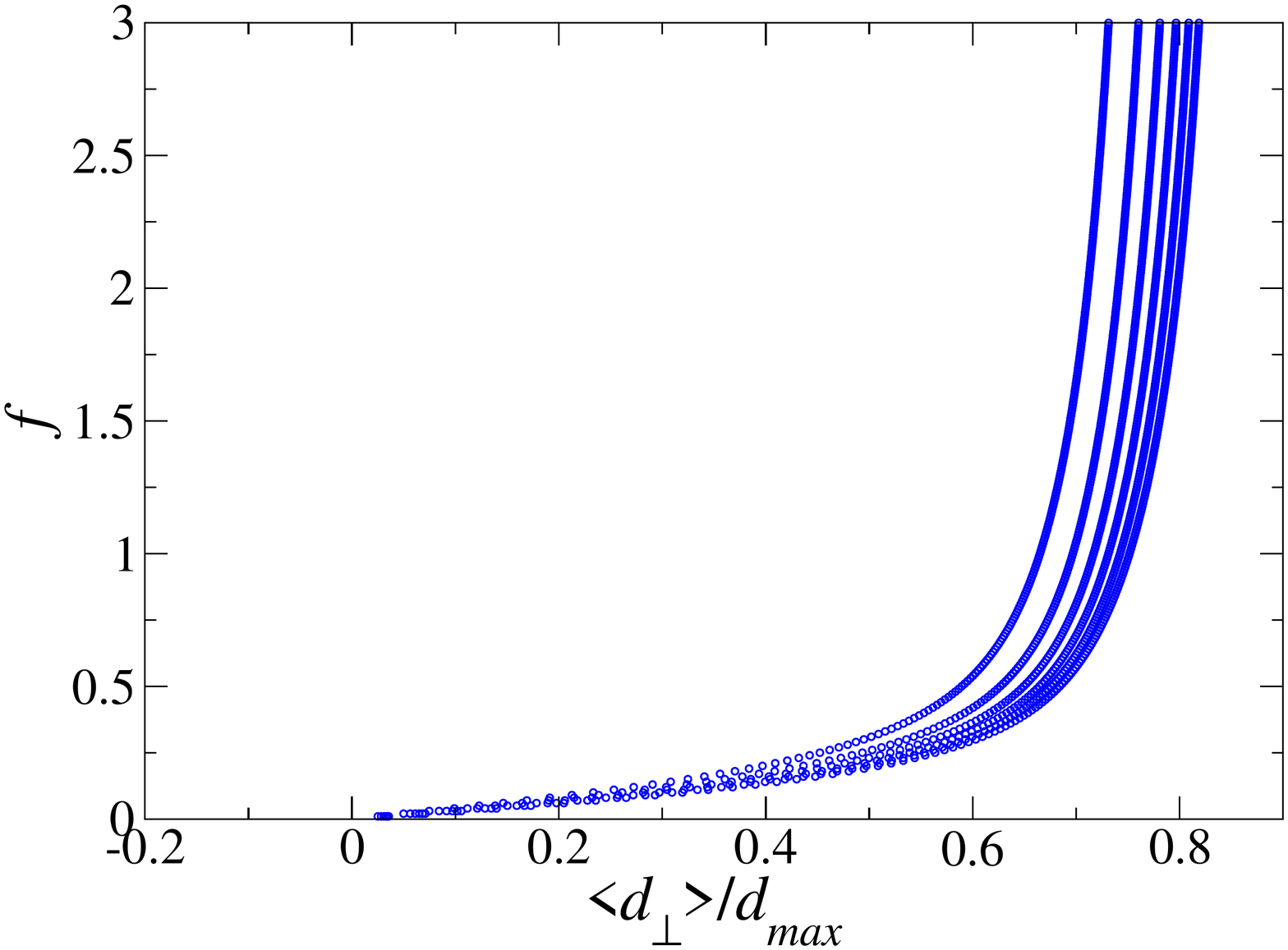}
	\end{center}\end{minipage} \hskip0.25cm	
\caption{Dependence of the average horizontal separation $\la d_\bot \ra$ (normalized to $d_{max}=l+2R$) on the 
external dimensionless force $f=\beta R F$ for different sets of parameters in the 
case of constrained symmetric wrapping $\alpha_1=\alpha_2 = 3\pi/8$. 
Left pane:  $\mu=10$ (high rigidity). Right pane: $\mu=5$  (low rigidity).
  Clearly, for this set of parameters only the extended phase is allowed for 
large values of force regardless of rigidity. Curves from left to right correspond to $l=nR$ for integer 
$n=3,\ldots,8$.}
\label{fig3}
\end{center}\end{figure*}

An interesting case with two cylinders was discussed in detail by Rudnick and 
Bruinsma \cite{rubr:1999}: they dealt with a system composed of two cylinders at 
a fixed arc-length separation $l$ and with fixed wrapping angles, 
solved in the Gaussian approximation level. We analyze here the exact solution 
for the two phases described in Ref. \cite{rubr:1999} which are (see also Fig. \ref{schematic2}):
\begin{itemize}
\item[(i)] the looped phase where the mean projected separation $\la d_\bot \ra$ 
is negative meaning that the cylinder at larger distance along the 
filament lies to the left of the other cylinder, thus causing the elastic filament  to 
loop and
\item[(ii)]  the extended phase where $\la d_\bot\ra$ is positive and there is no loop.
\end{itemize}
We fix $\alpha_1$ and $\alpha_2$, and choose a range of elastic filament  arc-lengths, $l$, between 
the cylinders. We calculate the average horizontal separation $\la d_\bot \ra$ for 
a given external tension $f$ (in dimensionless units) and plot $f$ versus $\la d_\bot \ra/(l+2R)$ for 
various choices of parameters. The average horizontal separation
is calculated from Eq. (\ref{dublerw}).

For symmetric wrapping described by $\alpha_1=\alpha_2=\pi$ and $\mu=10$ (high 
rigidity) the results are shown in Fig. \ref{fig1}, left pane. From left to right the lines 
are for $l=nR$ for integer $n=3\ldots 8$. For small forces the filament 
is in the looped phase. Increasing the force actually makes the loops larger as 
shown by $\la d_\bot \ra$ becoming more negative. Eventually the extended 
phase becomes preferable and we have a limit of full extension as the force 
increases. Small $l$ also makes the loops more energetically favorable. For 
$\alpha_1=\alpha_2=\pi$ but smaller (rescaled) rigidity, $\mu=5$ (Fig. \ref{fig1}, right pane), we see that only the extended phase 
is allowed for the largest values of $l$. This implies that for a 
given rigidity there is a minimum distance (and obviously maximum force) for loop formation.

The largest extension in both these cases is noticeably less than the hypothetical maximum $d_{max}= l+2R$. This can be understood for the chosen fixed wrapping angle of $\pi$ since 
high rigidity constrains the filament to be close to tangential at both the 
entry and exit points.

In Fig. \ref{fig3} we then plot $f$ versus $\la d_\bot \ra/d_{max}$ where $d_{max} = l+2R$ 
for $\mu=10$ (left) and $\mu= 5$ (right) with fixed $\alpha_1 = \alpha_2 =3\pi/8<\pi/2$. The curves for the two  rigidities are qualitatively similar with only the extended phase allowed. There is a 
small $l$ dependence, with curves corresponding to shorter $l$ lying to the 
right of those for longer $l$. Unlike in Fig. \ref{fig1}, the largest 
extension for both values of $\mu$ is now closer to $d_{max}$  since the fixed wrapping 
angle is smaller than $\pi$ and the constraint that the filament is tangential to 
the cylinders has a weaker effect. We observe that as the rigidity decreases the 
dimensionless displacement for a given (small) external tension also decreases; a lower rigidity 
allows the elastic filament  to fluctuate more and increase the entropic contribution to 
the free energy without incurring a large energy penalty. From our exact 
analysis, we thus qualitatively confirm the results for this system  by Rudnick and Bruinsma \cite{rubr:1999}. 

We will not discuss in detail the antisymmetric configuration $\alpha_2=-\alpha_1$
as here  a looped phase does not occur and the plots are similar to those shown in Fig. \ref{fig3}.

\begin{figure*}[t!]\begin{center}
	\begin{minipage}[b]{0.485\textwidth}
	\begin{center}
		\includegraphics[width=\textwidth]{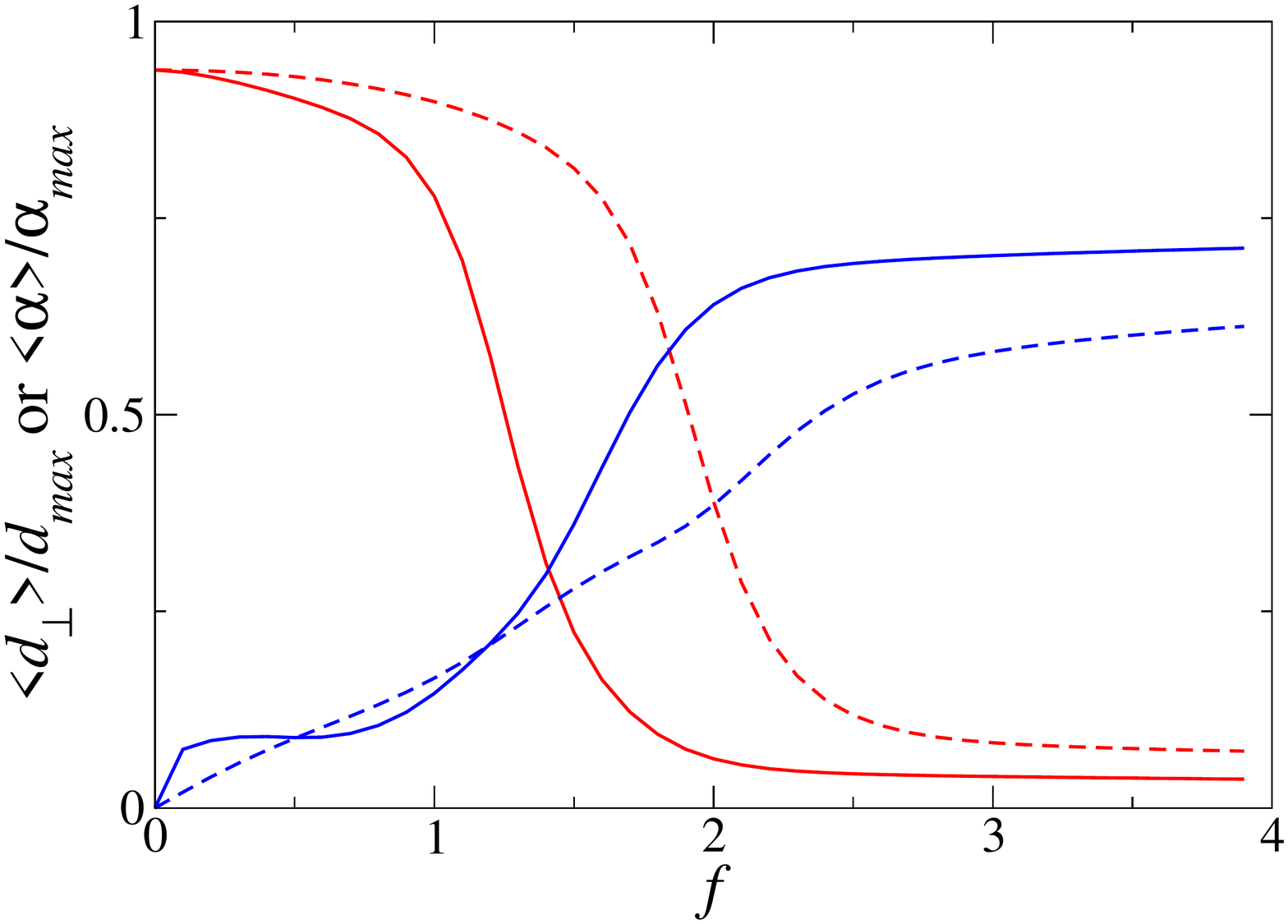}
	\end{center}\end{minipage} \hskip0.25cm
	\begin{minipage}[b]{0.49\textwidth}\begin{center}
		\includegraphics[width=\textwidth]{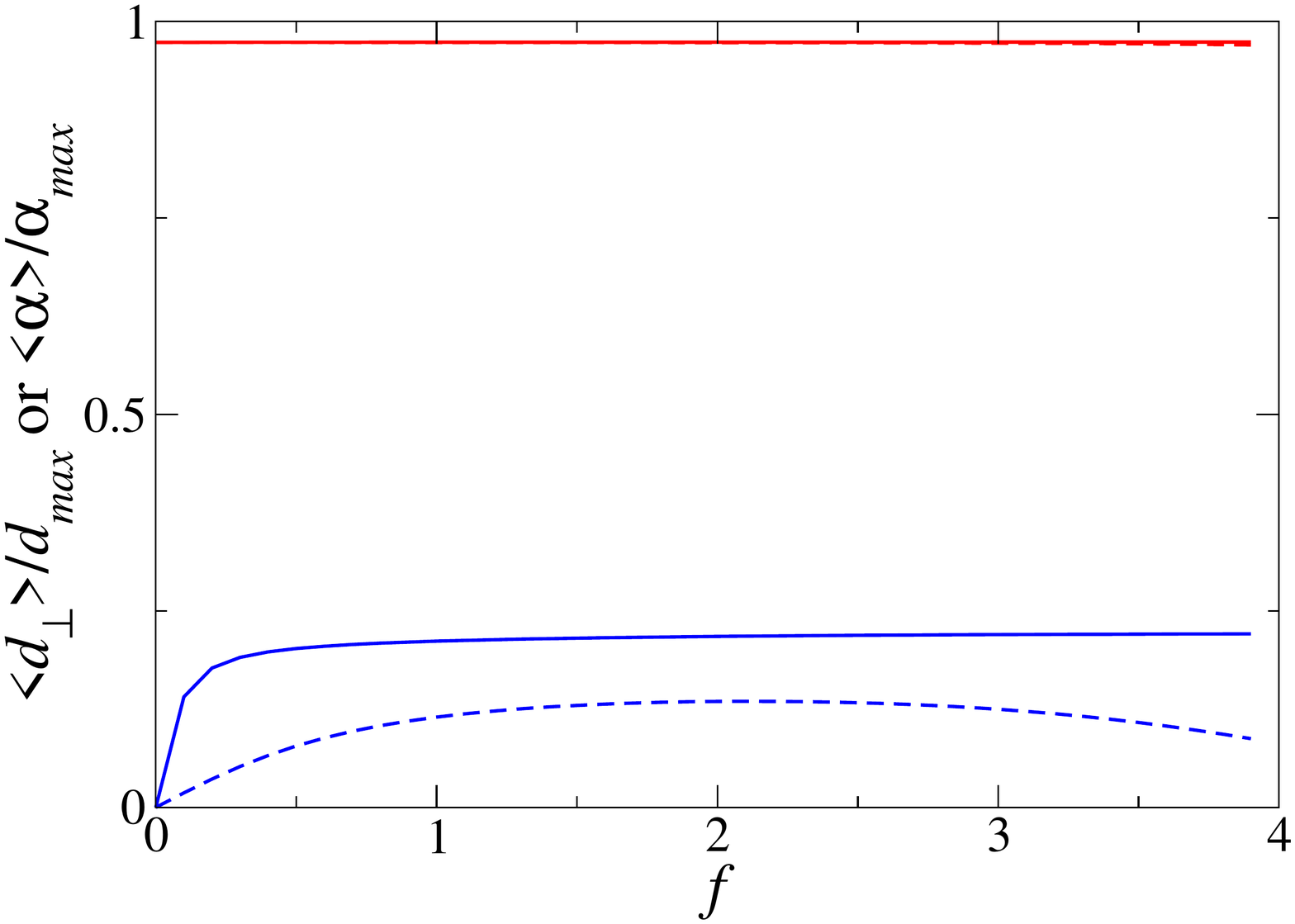}
	\end{center}\end{minipage} \hskip0.25cm	
\caption{Mean projected separation, $\la d_\bot \ra$ (blue curves), and mean wrapping angle, $\langle \alpha\rangle$ (red curves), for unconstrained 
wrapping of elastic filament  around two cylinders. Left pane: $\sigma=1$. Right pane: 
$\sigma=10$. 
Solid curves correspond to $\mu=10$ and dashed curves 
correspond to $\mu=1$. Note that the results have been normalized to maximum 
displacement, $d_{max}=l^\prime+2R$, 
or to maximum angle. }
\label{fig:alpha_sig1}
\end{center}\end{figure*}

\subsection{\label{two_unconstrained} Two cylinders: unconstrained wrapping}

We now allow the $\alpha_i$ to be a dynamical or annealed variable as opposed to the quenched case of the previous Section. As noted
in the beginning of Section \ref{unbinding} (case II), the 
corresponding realization would be an elastic filament  wrapped around two cylinders with 
a fixed arc-length separation $l^\prime$ between the two points of pinning and 
no other constraints. The length of elastic filament  adhering to the cylinders is now variable 
and so its length outside the region between the two cylinders 
and the effective length between them must both be allowed to vary 
dynamically.

In Section \ref{two_constrained} we gave an expression for the partition function, $Z(\alpha_1,\alpha_2,l)$,  
for fixed angles of wrapping and fixed effective length of elastic filament between the 
cylinders. Therefore the total 
partition function for dynamical wrapping is clearly given by
\begin{widetext}
\begin{equation}
Z=\int d\alpha_1 d\alpha_2 \int^{{\rm min}(l/R,\alpha_1)}_0 
d\alpha_1^\prime \int^{{\rm min}(l/R-\alpha_1^\prime,\alpha_2)}_0 d\alpha_2^\prime\;
Z(\alpha_1,\alpha_2,l=l^\prime-R\alpha_1^\prime-R\alpha_2^\prime),
\label{uncon_Z}
\end{equation}
\end{widetext}
where the integration is over the four wrapping angles which take into account 
the various ways for the wrapping to occur; for the $i$-th cylinder $\alpha_i$  
(or $\alpha_i^\prime$) is the total (or internal) wrapping angle of the elastic filament, where internal
refers to the length between the two cylinders. The computations were done with 
$\alpha_i, \alpha_i^\prime>0$; although the choice of sign for the first cylinder wrapping is arbitrary, 
we force symmetric wrapping on the second cylinder. We find that the antisymmetric configuration
gives similar results and so do not report on it in detail. $\la d_\bot \ra$ can now be 
calculated by averaging the expression in Eq. (\ref{dublerw}) over the four wrapping angles.
 

Because of the ability to pre-compute the various contributions to the partition 
function before any integration and/or summations are done, computation time is 
kept to a minimum. However, there is an increase of several orders of magnitude 
in the computing cost compared with the fixed wrapping case.

In the following computations, we chose $l^\prime=2\pi R$ and $\alpha_i <12\pi$. This puts
a reasonable limit on the maximum wrapping angles but it already has a significant
outcome and the amount of computing resources required is kept reasonable. 


In Fig. \ref{fig:alpha_sig1} we plot the average projected length $\la d_\bot 
\ra$ versus $f$ for $\sigma = 1,10$ and $\mu=1,10$. As $f$ is increased an 
unwrapping transition is indicated by a rapid decrease in $\la \alpha \ra$ and a 
corresponding increase in $\la d_\bot \ra$.  The results for $\sigma=1$ show a 
clear unwrapping transition (Fig. \ref{fig:alpha_sig1}, left). When the unwrapping occurs, the distance between 
the cylinders increases as one would expect, whilst in the wrapped phase there 
appears to be a small but potentially interesting effect causing a decrease in 
the separation as the force increases. This can be seen by the bump in the 
$\mu=10$ line at small $f$. As the rigidity $\mu$ is decreased the unwrapping is 
less pronounced; it is now easier for the filament between the cylinders to 
fluctuate more strongly on average.

For the case of $\sigma=10$ (Fig. \ref{fig:alpha_sig1}, right), there is no unwrapping 
transition for both values of $\mu$ over the range of forces indicated. One 
would expect this kind of outcome when one changes the binding energy $\sigma$. 
Higher binding energy prevents unwrapping, holding the cylinders tightly 
onto the  filament. The distance between the cylinders does however 
slowly increase with increasing rigidity owing to the straightening of the 
elastic filament.  

We also note, in this case, that, although $\alpha$ does not change (it 
stays maximally wound), the average projected length between the two cylinders 
decreases for large enough external force.  This counters the fact the 
filament will tend to straighten with increased external tension. It does, however, lead to the 
conclusion that for large external tension there is an effective 
attraction between the cylinders that pulls them together. This conclusion 
is corroborated also by direct evaluation of the interaction free energy between the 
cylinders as discussed below.

\subsection{\label{free_energy:absorption} Free energy of the unwrapping transition}

To understand the phase transition between a free and wrapped cylinder, i.e., 
the phenomenon of unwrapping and desorption, we calculate the free energy 
of wrapped and unwrapped systems. We restrict ourselves to the case of fixed wrapping angle (case I). This can be trivially calculated from our 
theory as we already have calculated the partition function:
\begin{equation}
\beta\Phi(\alpha_1,\alpha_2,l)=-\ln{Z(\alpha_1,\alpha_2,l)}.
\label{eq:Phi_1}
\end{equation}
Since we normalize all energies by subtracting the ground-state energy 
and do not put in wavefunctions for the end points, we normalize the system 
to have $Z=1$, $\Phi=0$ for a filament with no  wrapped cylinders.
\begin{figure*}[t!]\begin{center}
	\begin{minipage}[b]{0.485\textwidth}
	\begin{center}
		\includegraphics[width=\textwidth]{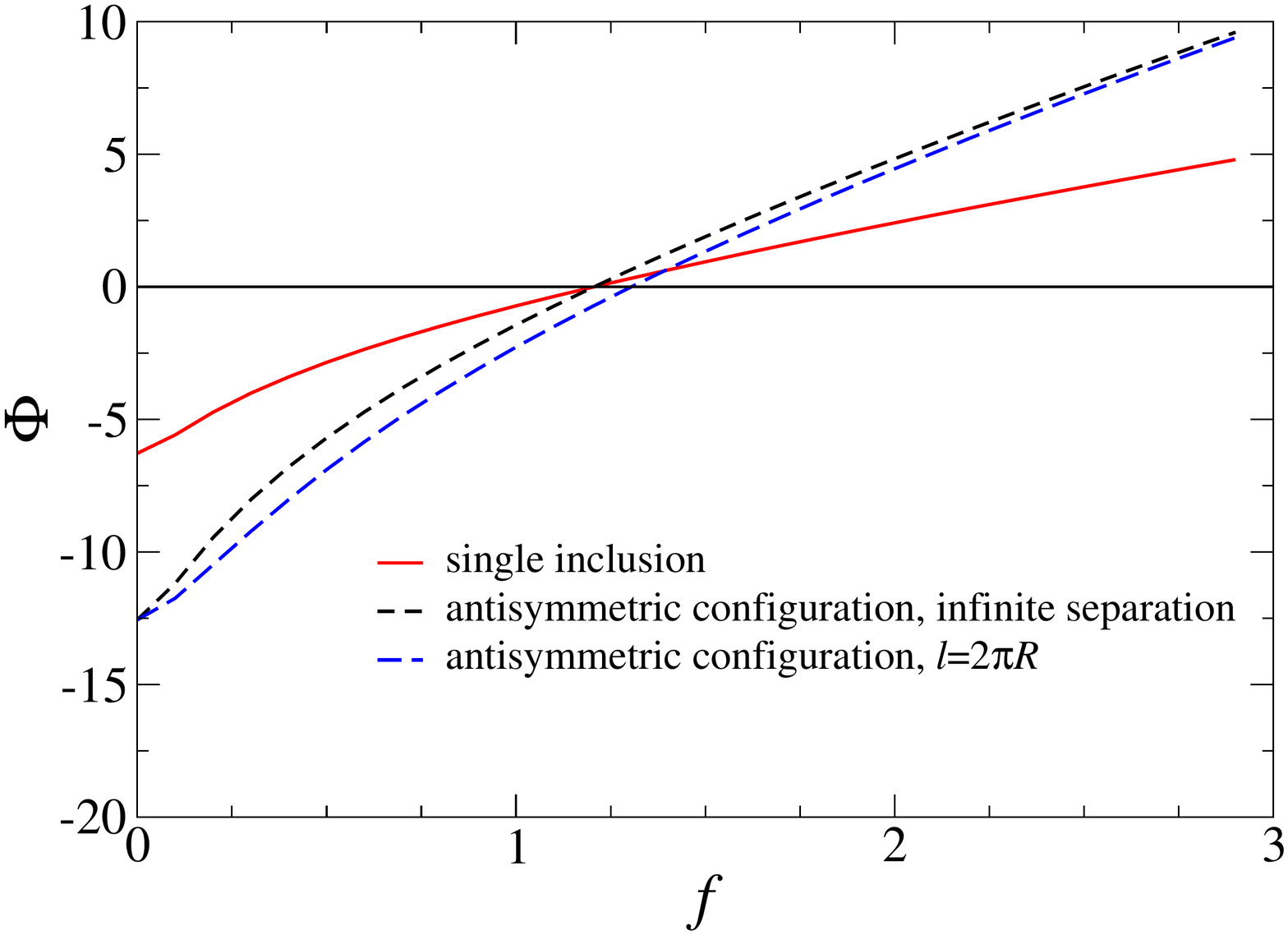}
	\end{center}\end{minipage} \hskip0.25cm
	\begin{minipage}[b]{0.49\textwidth}\begin{center}
		\includegraphics[width=\textwidth]{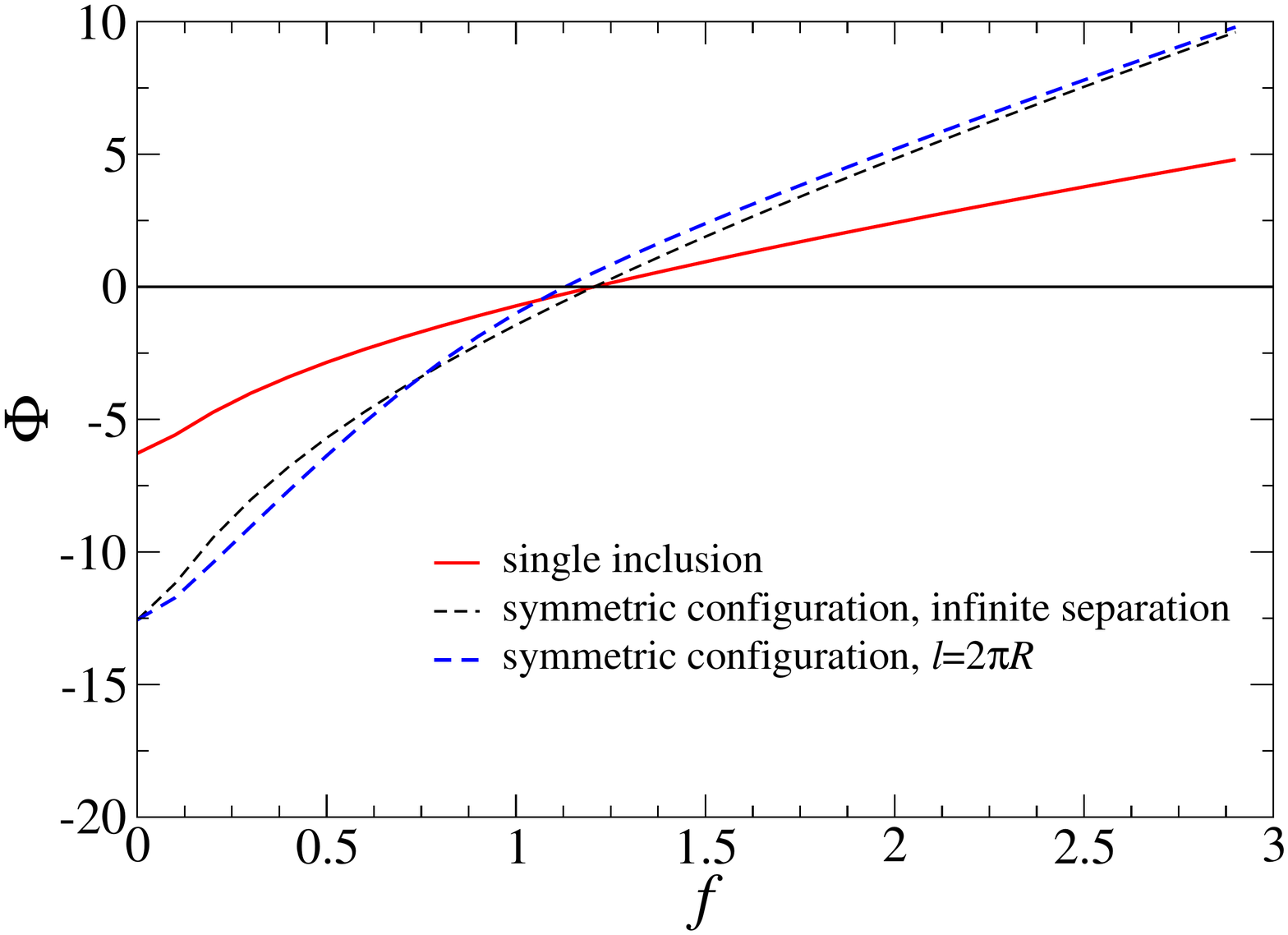}
	\end{center}\end{minipage} \hskip0.25cm	
\caption{Free energy as a function of the dimensionless external tension $f$ for $\mu=10$ and $\sigma=4.5$.
Left: Red solid curve corresponds to a single cylinder $\alpha=\pi$, dashed curves correspond to antisymmetric double 
cylinder, $\alpha_1 = -\alpha_2 = \pi$. The lower (blue) curve corresponds to $l=2\pi R$ and the upper (black) curve is limit of 
increased separation. Right: Red solid curve corresponds to a single cylinder $\alpha=\pi$, dashed curves correspond to symmetric double 
cylinder, $\alpha_1 = \alpha_2 = \pi$. The upper  (blue) curve corresponds to $l=2\pi R$ and 
the lower (black) curve is limit of increased separation. Note the double cylinder (black) curve that coincides with the
single cylinder (red) curve at free energy equal to zero gives the limit of infinite separation.}
\label{fig:antisymenergy}
\end{center}\end{figure*}
We consider the free energy of a system for the cases of a single wrapped cylinder with wrapping angle  
fixed to $\alpha=\pi$, and of the system of two wrapped cylinders in the symmetric and antisymmetric configurations
with $\alpha_1=\pi$ and $\alpha_2=\pm\pi$, respectively. In Fig. \ref{fig:antisymenergy} we show 
the free energy for these three cases. The lefthand and righthand plots compare the single cylinder with 
the double cylinder for the inter-cylinder separation $l = 2\pi R$ and the limit of large $l$ for the 
symmetric and antisymmetric configurations, respectively. We use  
$\mu=10.0$ and $\sigma=4.5$. For $f=0$, the free energy of the two 
cylinder system is essentially double that of a single cylinder, as would be 
expected. As the force increases so does the free energy.

In the antisymmetric case, at low external force the wrapping of both cylinders 
is energetically favorable. As the external tension  increases the two cylinders stay 
wrapped until both desorb simultaneously (the line crosses the x-axis). As we 
would expect in the limit of large separation, the critical force for 
unwrapping is the same as that of a single cylinder, however as the separation 
decreases a larger force is required as they are bound together.

In the symmetric case we also see that at low external tension the double cylinder 
wrapping is energetically favorable. However, as the tension increases there comes 
a point where the single cylinder wrapping becomes energetically preferred and 
one of the cylinders then unwraps and leaves the chain. The details of this 
process can not be captured appropriately in our model since we do not include 
the chemical potential for the unwrapped cylinders.
 
Increasing the external tension further unwraps the last cylinder. If the two cylinders are 
far apart we see the same behavior as for the antisymmetric case, as would 
indeed be expected. This agrees with previous studies showing that 
antisymmetric (symmetric) wrapping gives an attractive (repulsive) force \cite{rubr:1999}. 


\section{\label{effective_interaction} Effective interaction between two cylinders}

We now consider the problem of effective interaction mediated by the elastic filament's
fluctuations between two wrapped cylinders. The elastic and adhesive energy 
expressions are given as before.

\subsection{\label{constrained_wa} Constrained wrapping angles}

We first consider the case where the two cylinders are separated by fixed arc-length, $l$, 
and the wrapping angles are constrained at fixed values $\alpha_{1}$ and $\alpha_{2}$ (see the remarks identifying case I 
in Section \ref{unbinding} and \ref{two_constrained}). In numerical 
simulations we will assume furthermore that they are equal up to the sign, 
$\alpha_{1} = \pm \alpha_{2}$; the sign differentiates between the 
symmetric and the antisymmetric wrapping case. 
We can evaluate an effective interaction between the cylinders by 
fixing the (projected) horizontal separation $d_\bot$ between them and then 
calculating the corresponding free energy.  The projected separation in the 
embedding space is defined as
\begin{widetext}
\begin{equation}
d_\bot = R\,{\rm sgn}(\alpha_1)\sin(\psi_1+\alpha_1) + \int_{s_2}^{s_3} ds 
 \cos(\psi(s))-R\,{\rm sgn}(\alpha_2)\sin(\psi_2). \label{dbot}
\end{equation}
\end{widetext}
This constraint can be handled most conveniently by introducing an 
additional term into the total energy of the system  {\em via} a Lagrange multiplier of the 
form $\lambda d_\bot$,
where $\lambda$ can be interpreted as the force used to 
impose the constraint of fixed projected distance between the two cylinders in 
the horizontal direction.

The wrapping constraints can be 
implemented as before and one finds the appropriate partition function to be
\begin{widetext}
\begin{eqnarray}
&&Z_\lambda(\alpha_1, \alpha_2,l) = \int d\psi_1d\psi_2\, \langle 0|\hat{O}_\lambda(\psi_1,\psi_1+\alpha_1) 
e^{-H_\lambda l/R\mu}\hat{O}_\lambda(\psi_2,\psi_2+\alpha_2)|0\rangle, 
\end{eqnarray}
\label{lambda}
\end{widetext}
where
\begin{eqnarray}
H_\lambda\Psi_{\lambda m}(\psi)=\left(-\frac{d^2}{d\psi^2}-\mu (f-\lambda)\cos{(\psi)}\right)
\Psi_{\lambda m}(\psi) = \mu\varepsilon_{\lambda m}\Psi_{\lambda m}(\psi), 
\label{lambda1}
\end{eqnarray}
with
\begin{widetext}
\begin{equation}
\hat{O}_\lambda(\psi_1,\psi_1+\alpha_1) = \hat{O}(\psi_1,\psi_1+\alpha_1) \exp\left(-\beta \lambda R\,
{\rm sgn}(\alpha_1)\sin(\psi_1+\alpha_1)\right), 
\end{equation}
\end{widetext}
and
\begin{equation}
\hat{O}_\lambda(\psi_2,\psi_2+\alpha_2) = \hat{O}(\psi_2,\psi_2+\alpha_2) \exp\left(\beta \lambda R\,{\rm sgn}(\alpha_2)\sin(\psi_2)\right).
\end{equation}
With this constraint the partition function 
$Z({d_\bot})$ becomes
\begin{equation}
Z({d_\bot}, \alpha_1, \alpha_2,l) = \int_{C-i\infty}^{C+i\infty} \frac{{\mathrm d}\lambda}{2\pi i}\, 
e^{\beta  \lambda d_\bot}\, Z_{ \lambda}(\alpha_1, \alpha_2,l).  
\label{lagrange_mul}
\end{equation}
The appropriate constrained free energy then follows as
\begin{equation}
\beta\Omega({d_\bot}, \alpha_1, \alpha_2,l)=-\ln{Z({d_\bot}, \alpha_1, \alpha_2,l)}.
\end{equation}

The generalization to many wrapped cylinders is formally straightforward but 
computationally very tedious. One interesting future endeavour would be to assess 
the effects of non-pairwise additivity in the case of constrained and 
unconstrained wrapping around the interacting cylinders, i.e., the 
dependence of  effective two-cylinder interaction on the presence of other 
wrapped cylinders along the elastic filament.


\begin{figure*}[t!]\begin{center}
	\begin{minipage}[b]{0.485\textwidth}
	\begin{center}
		\includegraphics[width=\textwidth]{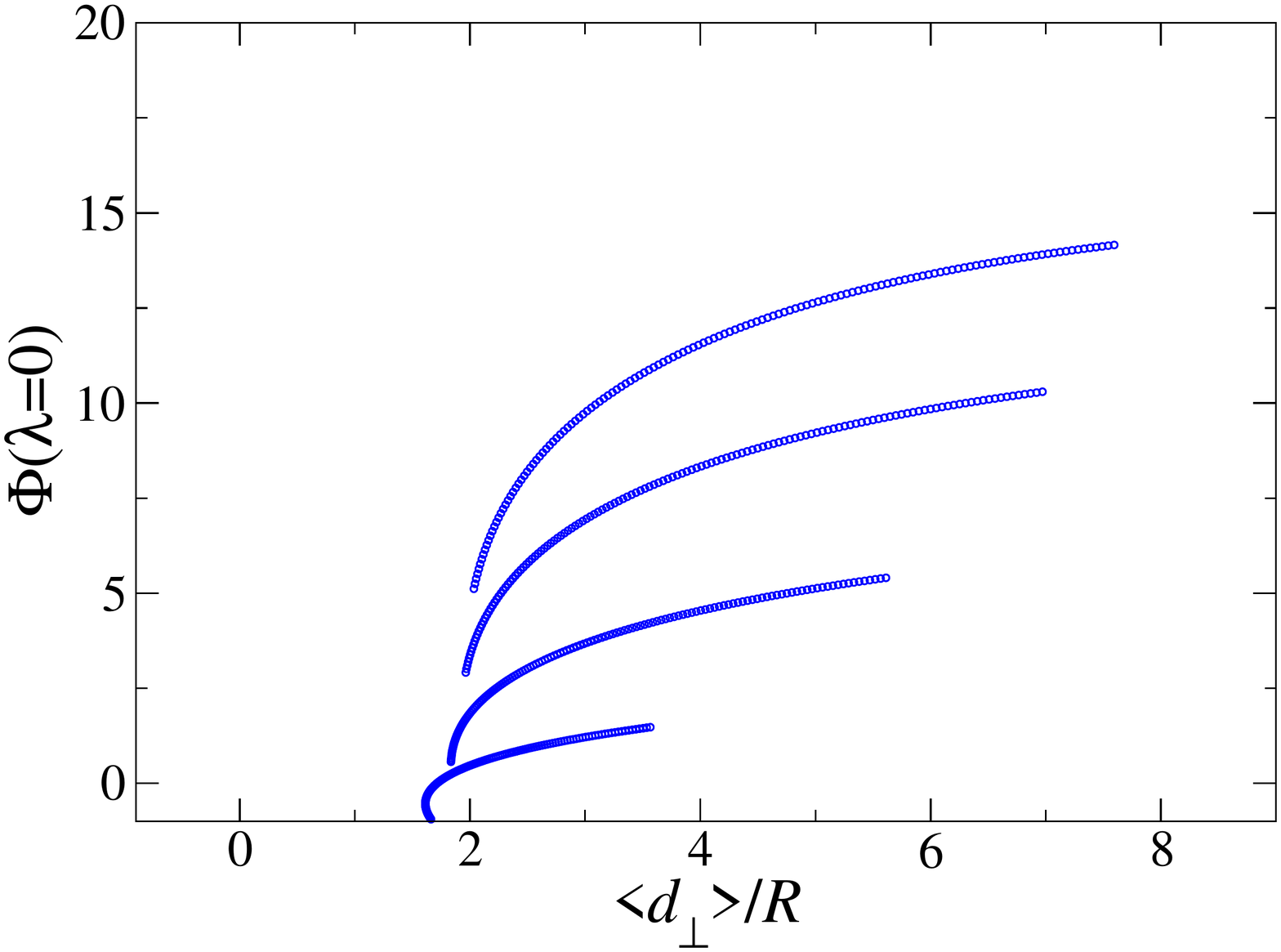}
	\end{center}\end{minipage} \hskip0.25cm
	\begin{minipage}[b]{0.49\textwidth}\begin{center}
		\includegraphics[width=\textwidth]{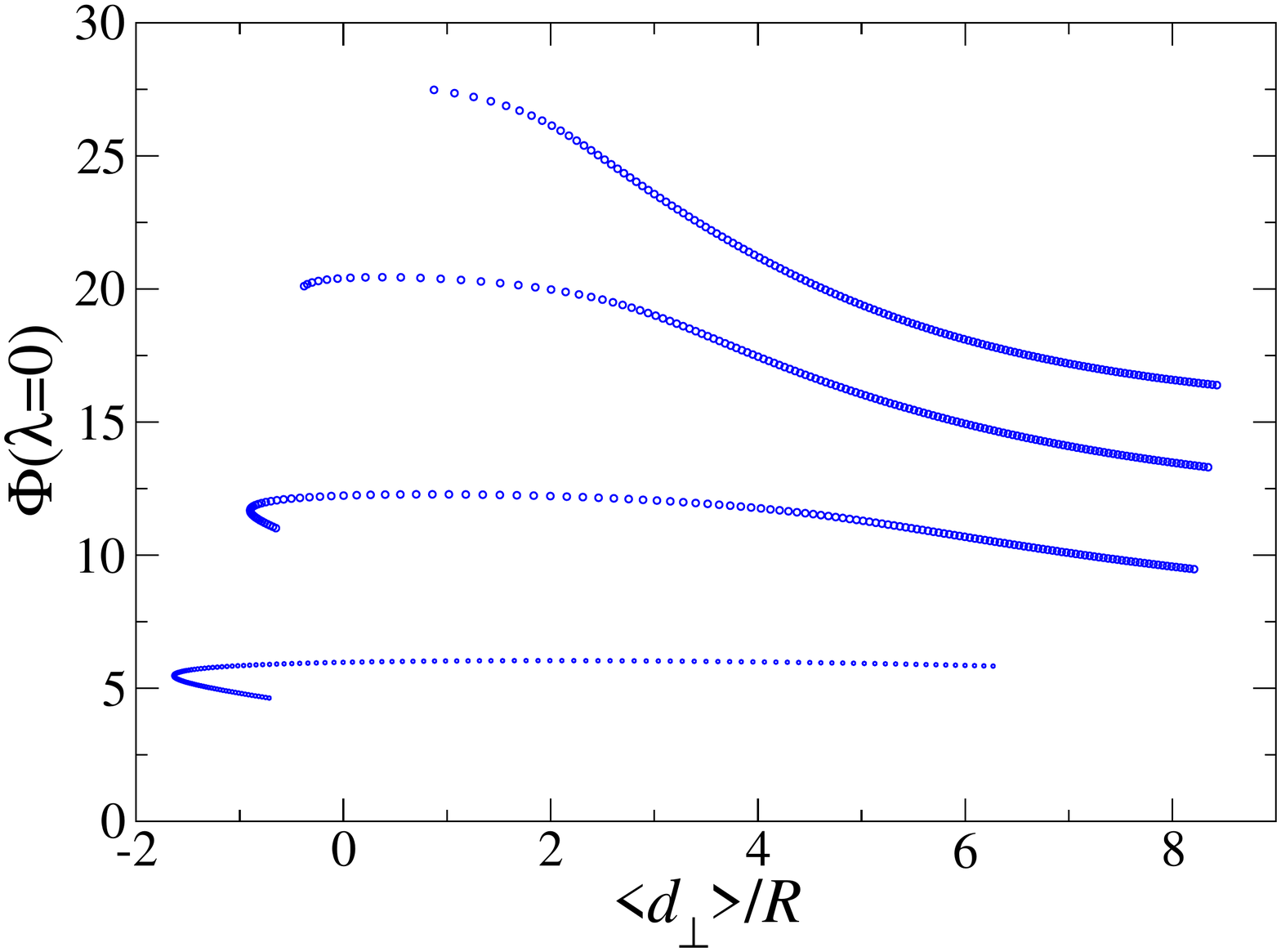}
	\end{center}\end{minipage} \hskip0.25cm	
\caption{Free energy as a function of the horizontal displacement projection 
$\la d_\bot\ra$ is shown for the  case of constrained wrapping with $\alpha_1 = \pm\alpha_2 = 
5\pi/8$,  $f = 0.4,1,2,3$ (from lower to upper curve), $\mu=50$ and $\sigma=13.0$. 
In the antisymmetric case (left) the effective interaction is attractive and in the symmetric case 
(right) it is repulsive.}
\label{fig:interactions}
\end{center}\end{figure*}

We can define related free energies appropriate for the sake of numerical calculation using
\begin{equation}
\beta\Phi(\alpha_1,\alpha_2,l,\lambda)=-\ln{Z_{\lambda}(\alpha_1, \alpha_2,l)}
\label{frencon}
\end{equation}
and its Legendre transformation
\begin{equation}
\Xi(l,\langle{d_\bot}\rangle,\alpha_1,\alpha_2)=\min_\lambda\left(\Phi(\alpha_1,\alpha_2,l,\lambda)-
\lambda\langle{d_\bot}\rangle\right),
\label{frencon2}
\end{equation}
which correspond to a system where either the polymer length, $l$, or the average horizontal displacement, 
$\langle d_\bot \rangle$, are fixed respectively. 

In Fig. \ref{fig:interactions} we plot $\Phi$ for $\lambda=0, \alpha_1 = 5\pi/8$ 
and $\alpha_2 = \pm \alpha_1$ corresponding to symmetric and antisymmetric 
configurations, respectively. 
The length of elastic filament connecting the cylinders, $l$, is not a dynamical variable in our current model. However, by repeating the simulation for a sufficiently large range of $l$ we are able to produce the curves for fixed external force $f = 0.4,1.0,2.0,3.0$ (from bottom to top). We assume that because there are pinning sites distributed along the filament, the cylinders can move through a tunnelling or hopping mechanism between sites and so $l$ and hence $\langle 
d_\bot \rangle(l)$ will vary to minimize the free energy $\Phi$. Thus we can infer from $\Phi$ the effective interaction between two cylinders in a sytem where $l$ is a dynamical variable.
We choose to plot $\Phi$ versus the value of 
$\langle d_\bot \rangle(l)$, the projected horizontal separation between the two 
cylinders, since this is the more relevant observable. One can see the looped 
phase and the extended phase, corresponding to negative and positive $\langle 
d_\bot \rangle$, respectively. In the extended phase $\Phi$ increases with the 
magnitude of the external force for both kinds of wrapping symmetry, as one would expect.
The effective interaction between the cylinders does, however, 
depend crucially on the symmetry of wrapping. In the asymmetric case (Fig. \ref{fig:interactions}, left pane) the effective interaction is attractive in the 
extended phase. In the symmetric 
case, however, (Fig. \ref{fig:interactions}, right pane), we see that the 
effective interaction is repulsive in the extended phase but then changes sign 
in the looped phase. In the extended phase, therefore, the cylinders with symmetric 
(antisymmetric) wrapping will move towards smaller (larger) $\langle d_\bot 
\rangle$; there is an effective attractive (repulsive) force between the 
cylinders. The higher the external force the bigger this effective 
force is between the wrapped cylinders.

\begin{figure*}[t!]\begin{center}
	\begin{minipage}[b]{0.485\textwidth}
	\begin{center}
		\includegraphics[width=\textwidth]{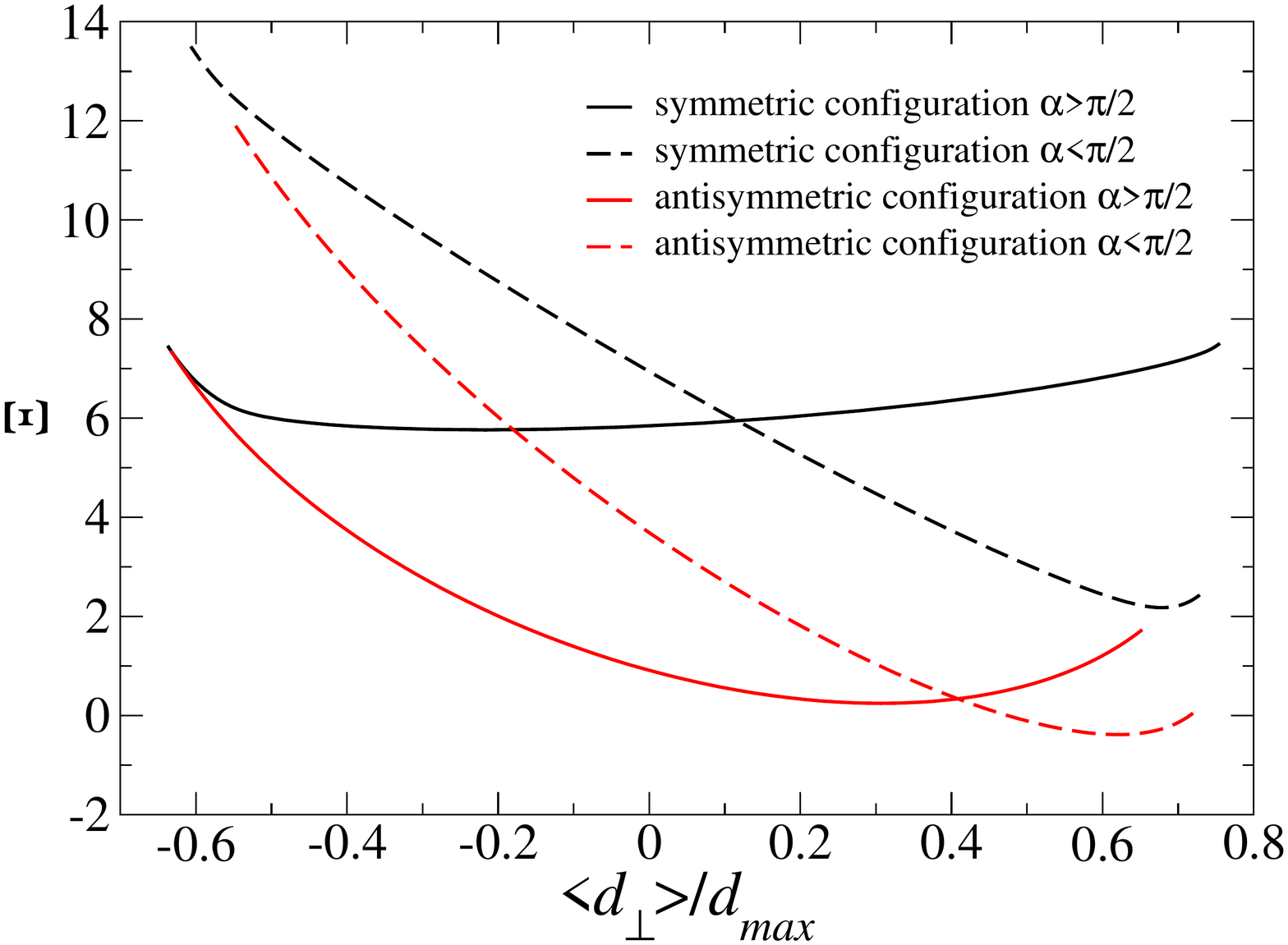}
	\end{center}\end{minipage} \hskip0.25cm
	\begin{minipage}[b]{0.49\textwidth}\begin{center}
		\includegraphics[width=\textwidth]{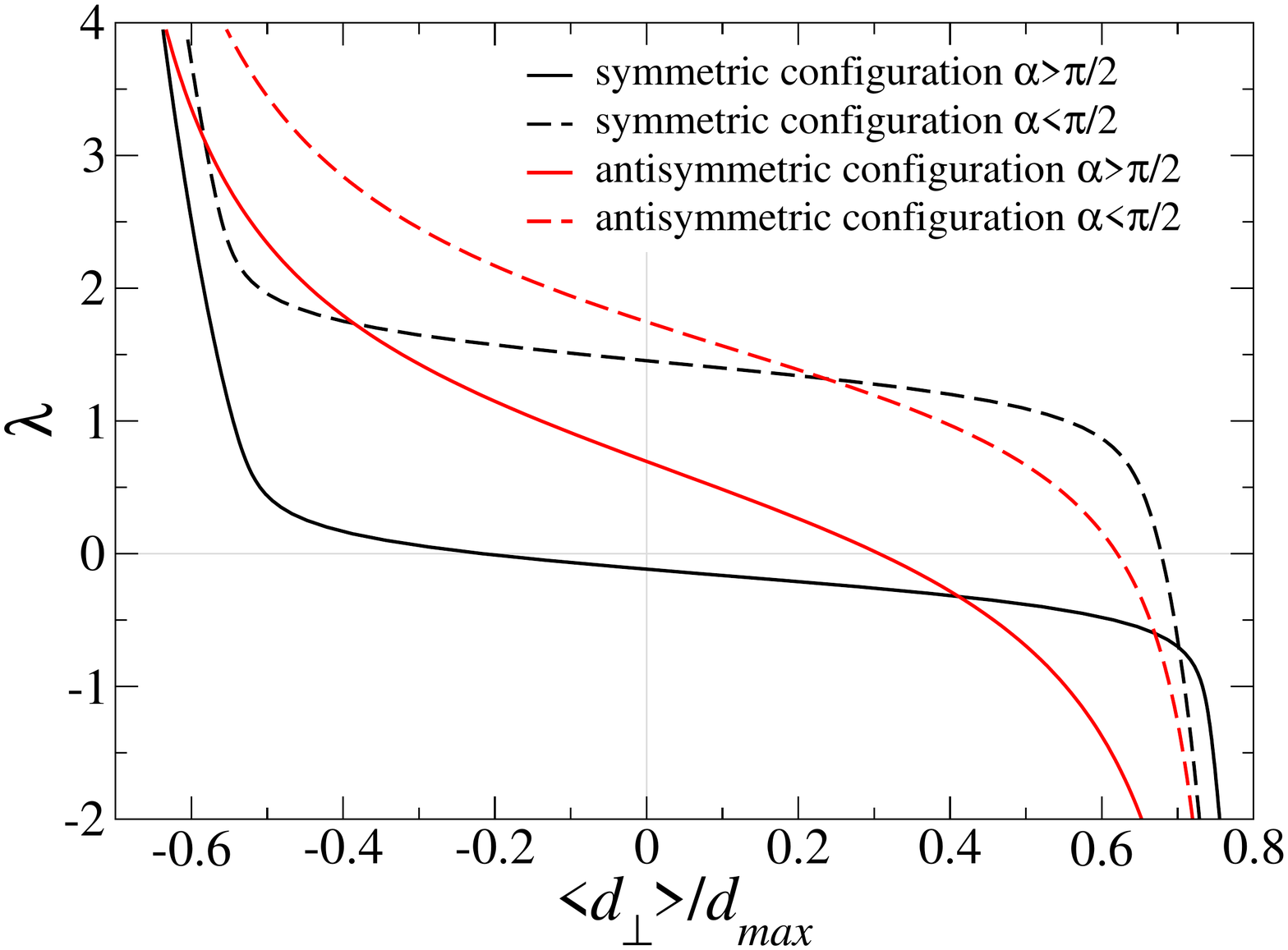}
	\end{center}\end{minipage} \hskip0.25cm	
\caption{Left: the free energy as a function of the horizontal displacement 
projection $\la d_\bot\ra$ (normalized to $d_{max}=l+2R$). Right: the required 
externally applied force, $\lambda$, as a function of the horizontal 
displacement projection $\la d_\bot\ra$. Here we consider the case of constant wrapping 
with $\alpha= 5\pi/8>\pi/2$ and $\alpha= 3\pi/8<\pi/2$ for $f=0.4$, $\mu=50$, $\sigma=13.0$ and $l=4\pi R$.}
\label{fig:interactions2}
\end{center}\end{figure*}

In Fig. \ref{fig:interactions2} we plot the free energy $\Xi(l,\langle d_\bot \rangle,\alpha_1,
\alpha_2)$  defined in Eq. (\ref{frencon2}); $\Xi$ is a function of $\langle d_\bot \rangle$
at fixed $l$ (i.e.,  $l=4\pi R$ in the figure). 
 In the lefthand plot we show $\Xi$ for both symmetric and antisymmetric 
wrappings for $f=0.4,\mu=50,\sigma=13.0$ and $\alpha = 5\pi/8, 3\pi/8$. The significance of
the values chosen for $\alpha$ is that they are, respectively, $\gtrless \pi/2$. For 
symmetric wrapping the equilibrium (minimum of $\Xi$) is in the looped phase for
$\alpha < \pi/2$ but it is in the extended phase for $\alpha > \pi/2$. In contrast for
antisymmetric wrapping, equilibrium is in the extended phase for both wrapping 
angles although at a smaller value of $\langle d_\bot \rangle(l)$ for $\alpha < \pi/2$
than for $\alpha > \pi/2$. In the righthand plot we show the projected separation 
$\langle d_\bot \rangle$ versus its conjugate variable $\lambda$. From Eq. (\ref{frencon2})
we note that $\lambda$ is determined as a function of $\langle d_\bot \rangle$ via 
\begin{equation}
 \lambda =  -\frac{\partial \Xi}{\partial \langle d_\bot \rangle}.
\end{equation}
In an extended phase, $\langle d_\bot \rangle > 0$, we have that
$\lambda > 0$ (or $\lambda < 0$) corresponds to an external force applied between the 
cylinders with magnitude $|\lambda|$ which pushes them together (or pulls them apart). In 
a looped phase, $\langle d_\bot \rangle<0$, we have that $\lambda > 0$ (or $\lambda < 0$) 
corresponds to an external force applied between the cylinders with magnitude $|\lambda|$ 
which pulls them apart (or pushes them together). We conclude that 
$\lambda\langle d_\bot \rangle > 0$ (or $\lambda\langle d_\bot \rangle < 0$) 
corresponds to an attractive (or repulsive) force between the cylinders. 

These four choices for the variables $\langle d_\bot \rangle$ and $\lambda$ are shown
separated by the dotted lines in the plot.
The equilibria of the lefthand plot in Fig. \ref{fig:interactions2} correspond 
to $\lambda = 0$ in the righthand plot. Clearly, for symmetric wrapping and a wrapping 
angle $\alpha \sim \pi/2$ the equilibrium value of the projected length is 
$\langle d_\bot \rangle \sim 0$ giving a high probability for the cylinders
to interact. 

One should not forget here that in reality there may be other interactions between the 
wrapped cylinders (as in nucleosomes) that are not taken into account in this model: there would be 
direct electrostatic repulsions acting in real 3D space, as well as short 
range steric interactions when the cylinders are wrapped symmetrically, but not 
when they are wrapped antisymmetrically. It is the sum of all these complicated 
interactions that would need to be taken into account in a complete theory of 
cylinder wrapping.

\subsection{Unconstrained wrapping angles}

Next we consider the case of two cylinders with no constraints on the wrapping 
angles, corresponding to the second case (case II) as explained in the beginning of 
 Section \ref{unbinding} and \ref{two_unconstrained}. 
 This case is more complicated from the previous one as it entails an 
additional integration with respect to the two wrapping angles. The partition 
function can then be written in the same way as before in Eq. (\ref{uncon_Z})  
{\bacol
\begin{equation}
Z_{\lambda}(l^\prime)=\!\int d\alpha_1 d\alpha_2 \int^{{\rm min}(l/R,\alpha_1)}_0 \!
d\alpha_1^\prime \int^{{\rm min}(l/R-\alpha_1^\prime,\alpha_2)}_0 \!d\alpha_2^\prime\;
Z_\lambda(\alpha_1,\alpha_2,l=l^\prime-R\alpha_1^\prime-R\alpha_2^\prime),
\label{uncon_Z0}
\end{equation}}
and the variables are defined in Section \ref{two_unconstrained} following 
Eq. (\ref{uncon_Z}).
We consider the free energy, now for unconstrained wrapping angles, given by
\begin{equation}
\beta\Phi(l^\prime,\lambda)=-\ln{Z_{\lambda}(l^\prime)}.
\label{frencon3}
\end{equation}
and its Legendre transformation
\begin{equation}
\Xi(l^\prime,\langle{d_\bot}\rangle)=\min_\lambda\left(\Phi(l^\prime,\lambda)-
\lambda\langle{d_\bot}\rangle\right),
\label{frencon4}
\end{equation}
which again correspond to a system where either the polymer length, $l^\prime$, or the average horizontal displacement, 
$\langle d_\bot \rangle$, are fixed respectively. 


\begin{figure*}[t!]\begin{center}
	\begin{minipage}[b]{0.485\textwidth}
	\begin{center}
		\includegraphics[width=\textwidth]{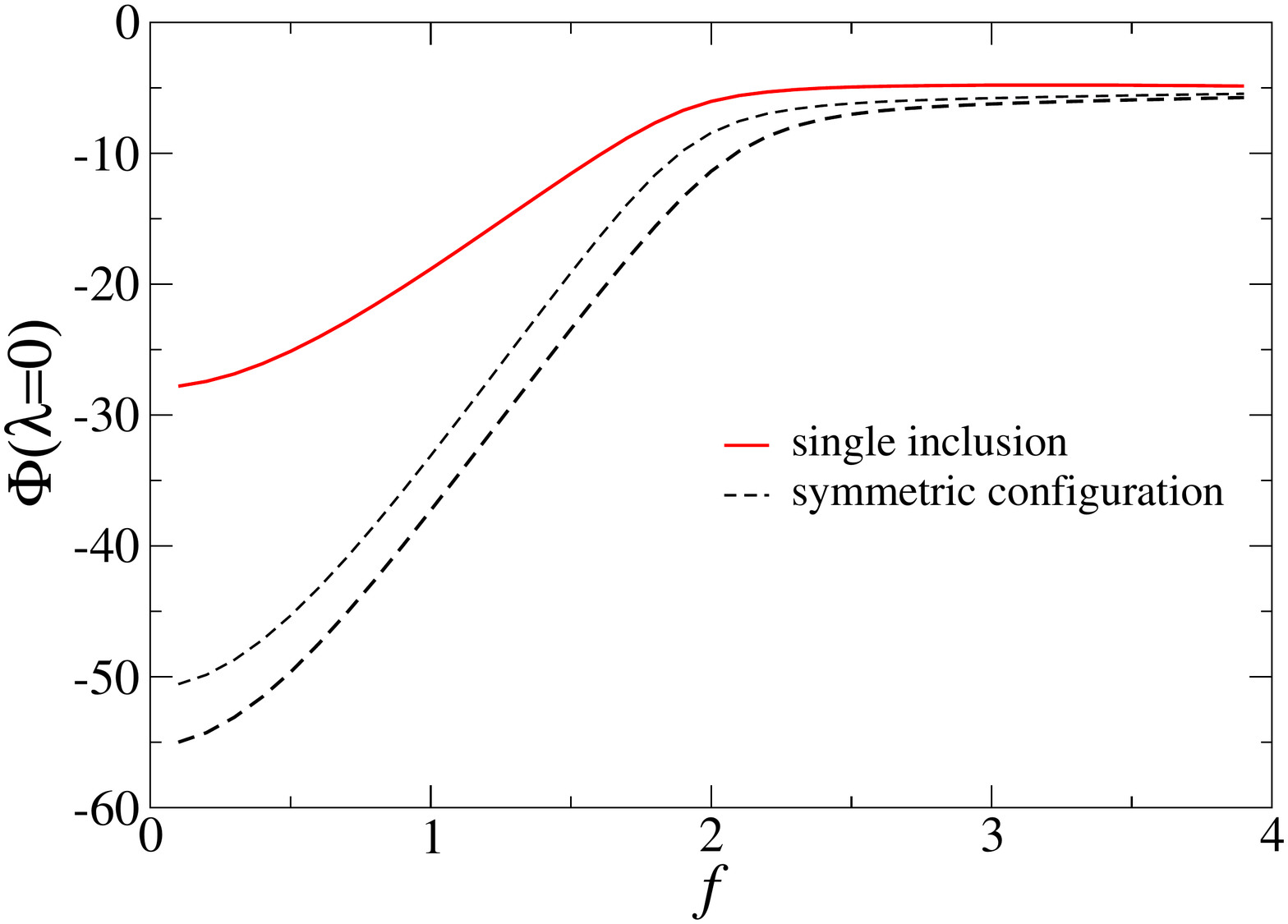}
	\end{center}\end{minipage} \hskip0.25cm%
	\begin{minipage}[b]{0.49\textwidth}\begin{center}
		\includegraphics[width=\textwidth]{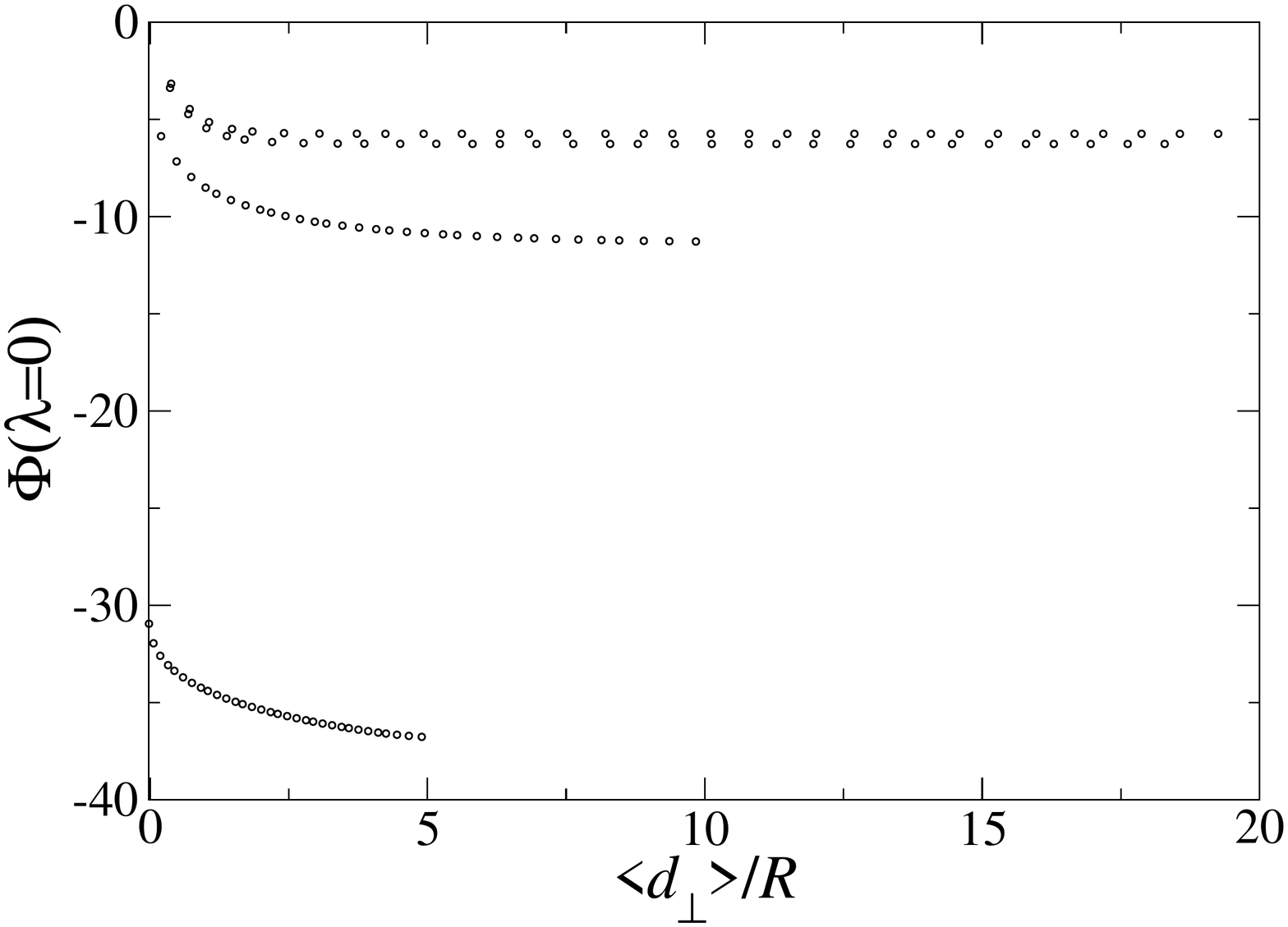}
	\end{center}\end{minipage} \hskip0.25cm	
\caption{The free energy for the case of unconstrained wrapping angles. Left: the free energy 
is shown as a function of external force $f$ for a 
single cylinder and for two cylinders with symmetric wrapping. 
The upper dashed 
curve is for $l'=\pi R$ and the lower dashed curve corresponds to $l'=20 \pi R$, being effectively 
infinite. As one expects, for large separation and low force, the free energy of two 
cylinders is double that of a single cylinder (red curve).  Right: the free energy as a 
function of projected horizontal distance $d_\bot$, for $f=0.01,1,2,3$ 
(from bottom to top). The free energy  decreases with separation (entropic effect) and increases with force. 
In both cases $\mu=1$ and $\sigma=1.25$.}
\label{fig:interactionsC}
\end{center}\end{figure*}

In the lefthand pane of Fig. \ref{fig:interactionsC} we plot the free 
energy, $\Phi$, of a quenched $l^\prime$ system (see remarks for case II  in the beginning of 
 Section \ref{unbinding}) as a function of external force $f$ for 
a single cylinder and for two cylinders with symmetric wrapping and various arc-length separations. For antisymmetric wrapping the results are almost identical 
and we do not show them separately. We see that as $f \rightarrow 0$ and the 
arc-length separation $l^\prime$ becomes very large ($\langle d_\bot \rangle \to \infty$), 
the free energy of two cylinders is 
twice the free energy of a single cylinder, which is a good consistency check. 
Note that for large $f$ the elastic filament can unwrap which allows the energetically 
favored small wrapping angle: The cylinders are pinned with no significant length
of filament  wrapped on them. This is in strong contrast to the case of constrained
wrapping angle shown in Fig. \ref{fig:antisymenergy} which shows a desorption 
transition due to the constraint imposed on the wrapping angle. 
In the righthand pane, we show the dependence of the unconstrained 
free energy, $\Phi$, on the projected separation $\langle d_\bot\rangle$. Although 
$l^\prime$ is a quenched variable, we are again able to produce these curve by 
repeating the numerical calculation for a sufficiently large range of $l^\prime$. The cylinders can move 
through a tunnelling or hopping mechanism between pinning sites and so $l^\prime$ and hence $\langle 
d_\bot \rangle(l^\prime)$ will vary to minimize the free energy $\Phi$. As the projected separation 
between the cylinders becomes smaller, i.e. they get closer together, 
we get an increase in the free energy corresponding to effective repulsive interactions which are not due to any hard-core repulsion between the cylinders, but are entropically generated. 
This entropy stems primarily from an ``entropic wrapping" effect: By limiting the space between the cylinders 
they cannot wrap in as many ways as for large separations. Such entropic 
wrapping effects should be distinguished from the usual entropic configuration 
effects in semi-flexible polymers. In the latter the number of configurations of the 
polymer chain changes as we restrict the position of its ends and this leads to 
entropic polymer elasticity; for entropic wrapping effect the physical picture is altogether different.

In Fig. \ref{fig:interactions3} we plot the conjugate variable $\lambda$ against the 
horizontal displacement $\langle d_\bot \rangle$ for a system with $l^\prime=2\pi R$. 
Again $\lambda$ can be interpreted as an external force needed to maintain a mean 
projected separation $\langle d_\bot \rangle$, as described in Section \ref{constrained_wa}. 
The response of the system to the external force is monotonic and attests to the fact  that a large repulsive 
force, $\lambda>4$, greater than the external applied force, $ f=3$ in this case, would be required to 
sustain a looped phase. The choice $l^\prime=2\pi R$ is relatively short and we do not include the exclusion 
of one cylinder by the other which would restrict the possible configurations especially in a looped phase. 
However, we treat the result here as an idealized case which is indicative of the possible likely 
configurations; in practice, the cylinders are not of infinite extent and so can be assumed to pass 
by each other more readily than the cylinders of this model. The absence of a looped phase in this idealized case is
strong evidence that it is not likely to occur except for large $\lambda$ in a more realistic
model of unconstrained wrapping. It should also be noted that the amount of computation time
to explore all the parameter space in this case is considerable and so we are confined to
investigating whether or not there is any significant non-trivial configuration that is
likely to be realized. 

We do not plot an analogue of the lefthand pane in Fig. \ref{fig:interactions2} as it does not 
provide any extra information, however the free energy, $\Xi$, can be easily 
calculated numerically in our formalism.

\begin{figure*}[t!]\begin{center}
	\begin{minipage}[b]{0.49\textwidth}\begin{center}
		\includegraphics[width=\textwidth]{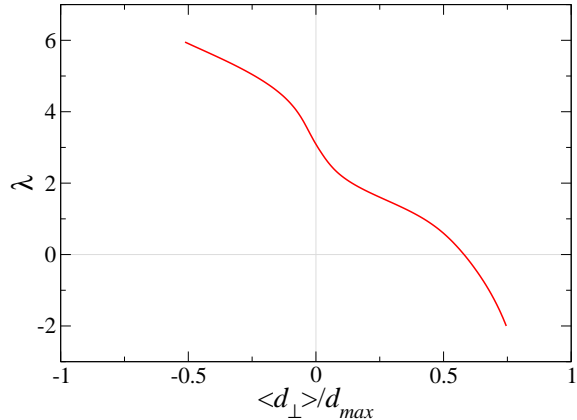}
	\end{center}\end{minipage}
\caption{The required externally applied force, $\lambda$, as a function of the 
horizontal displacement projection $\la d_\bot\ra$ (normalized to $d_{max}=l+2R$) is shown for the
 case of unconstrained wrapping with $f=3$, $\mu=1$, $\sigma=1.25$ and $l^\prime=2\pi R$. }
\label{fig:interactions3}
\end{center}\end{figure*}

From the numerical solution presented above it thus follows that the cylinder 
wrapping and the associated entropy presents yet another, apparently more 
important, source of polymer-mediated interactions between wrapped cylinders. 
Its source is the wrapping degrees of freedom that are constrained as the 
cylinders move closer together. To our knowledge, this source of effective 
interactions along a polymer chain has not before been clearly discussed in the  literature.


\section{Summary and Conclusions}
\label{sec:sum}

We have analyzed wrapping equilibria of one and two cylinders on a semi-flexible filament
driven solely by the elastic energy of the filament and the (adhesive) energy of wrapping  
around the cylinders. We derived the statistical properties and the free energy 
of wrapping in the case of one and two cylinders as well as the effective 
interaction free energy between two wrapped cylinders along the elastic 
filament. Our calculation is based on the functional integral representation of 
the partition function for the filament  and is exact, within the confines 
of the worm-like chain model, the assumed form of the wrapping potential and the 
limit of a 2D Eulerian plane.

In Section \ref{theory} we presented the generalized theory for elastic filament wrapping on one or 
more cylindrical cylinders in one dimension. 
In Section \ref{one_wa} we calculated the wrapping transition 
for a single cylinder and showed that it is necessary to solve the full 
Schr\" dinger  equation in order to obtain a good numerical value for the critical unwrapping external force. 
In Section \ref{unbinding} we analyzed the exact solution for two cylinders pinned a fixed
length apart on the elastic filament for both the looped phase, where the 
mean projected separation $d_\bot$ is negative, and for the extended phase, where $d_\bot$ is 
strictly positive and there are consequently no loops. The two cases considered are of 
constrained and unconstrained wrapping, respectively. In Section \ref{two_constrained} the 
wrapping angles $\alpha_1$ and $\alpha_2$ on the respective cylinders are fixed: the case of 
constrained symmetric wrapping. This gives the most interesting results concerning the
presence of a looped phase as discussed already by Rudnick and Bruinsma \cite{rubr:1999}.  
We considered two values of wrapping angles: $\alpha_1 = \alpha_2 = \pi$ and
$\alpha_1 = \alpha_2 = 3\pi/8$. For the larger value there is a clear looped phase
shown in Fig. \ref{fig1} characterized by the average horizontal separation $\langle d_\bot \rangle < 0$.
The loop initially increases in size as the external tension $f$ is increased and
eventually for sufficiently large $f$ the system switches over  to the extended phase. In contrast, for 
smaller value of $\alpha_i$ shown in Fig. \ref{fig3} there is no looped phase.   
In both cases the largest possible extension is close to its maximum possible value, $d_{max}$ as determined by the arc-length of filament between the two cylinders and their radii, as one would 
expect. We conclude that a looped phase is possible for constrained symmetric wrapping and 
sufficiently large wrapping angles but is absent if the wrapping angle is too small.
From Fig. \ref{fig1} we see that where it does occur, the maximum loop size increases as the 
rigidity $\mu$ increases and is sustained for a range of tensions $f$; for $\mu=10$
the most negative values of $\langle d_\bot \rangle$ are approximately for $0.2 < f < 1$.

In Section \ref{two_unconstrained} the two cylinders are pinned a distance $l^\prime$
apart along the contour of the elastic filament and the wrapping angles $\alpha_i$ are 
now dynamical (annealed) variables. In contrast to the
constrained case the wrapping energy encoded in the dimensionless variable $\sigma$ plays 
a direct role in the values of the observables. The wrapping angle on a given cylinder is
divided into two parts, which are the wrapping angles of the elastic filament wrapped to the left and 
to the right of the pinning point. In the two-cylinder case there are then four dynamical angle
variables over which to sum, and this greatly increases the computer time required to 
carry out the calculation. We discussed in detail only the symmetric wrapping configuration as the
antisymmetric one gives similar results. In Fig. \ref{fig:alpha_sig1} we show both $\langle d_\bot \rangle$ 
and $\langle \alpha\rangle$ (normalized to their maximum values) as a function of $f$ for $\sigma=1,10$ 
corresponding to small and large wrapping energy. For $\sigma=1$ there is a clear unwrapping transition 
for the two values of rigidity $\mu=1,10$ as the external tension $f$ increases, but for larger value of
$\sigma$ the system remains maximally wound for all values of $f$. In particular, there is no
looped phase indicated by $\langle d_\bot \rangle < 0$; this is characteristic of the unconstrained model.
We chose separation $l^\prime = 2\pi R$ which is relatively short compared with the cylinder radius $R$ and 
treat the results as an idealized case 
which is indicative of the possible more realistic configurations. In practice, the cylinders are not cylindrical 
or of infinite lateral extent and so can be assumed to pass by each other more readily than the cylinders 
in the present model. 

In Section \ref{free_energy:absorption} we calculate the free energy  of the symmetric 
and antisymmetric constrained systems of two cylinders pinned a distance $l$ apart with wrapping angles
$\alpha_1=\pm\alpha_2=\pi$, respectively, and compare with the free energy of a single pinned 
cylinder with $\alpha=\pi$. In all cases the free energy is normalized by subtracting that of the 
elastic filament with no pinned cylinders. 
The results show that for large $l$ 
and for both the symmetric and antisymmetric configurations, we observe that desorption 
occurs for single as well as double cylinder systems simultaneously
and for the same external tension. For small $l$ the situation is however different. For the antisymmetric configuration
the double cylinder system is more strongly bound than for a single cylinder and remains bound 
after the single cylinder has been desorbed. In contrast, for the symmetric configuration, as the force 
increases, first the double cylinder becomes unstable leading to a single cylinder desorption, leaving a bound single
cylinder which then desorbs as the external tension is increased further. The conclusion is that
the symmetry of the double cylinder constrained wrapping has a crucial effect on the desorption
transition. For unconstrained wrapping there is little structure since we do not associate a significant
binding energy with the pinning site itself and so, as the external tension increases, the cylinders simply unwrap but
remain pinned nevertheless. The effect in the constrained case is due to the competition between the wrapping and
the entropic contributions to the free energy as a function of external traction and separation between the cylinders.        

In Section \ref{effective_interaction} we investigated the induced force between two pinned
cylinders by introducing a force $\lambda$ conjugate to the projected distance $d_\bot$. 
We conclude that in the case of constrained wrapping the effective force, given by the slope of
the curves, depends on the symmetry of configuration, being {\em repulsive} in the symmetric case and 
{\em attractive} in the antisymmetric case. The dependence of the effective force on the arc-length 
separation $l$ between the cylinders follows closely the dependence on the
projected separation between the cylinders in the direction of the external
tension. In contrast, for the case of unconstrained wrapping we observe repulsion 
both for the symmetric as well as antisymmetric configurations. We interpret this repulsion 
as due to {\em wrapping entropy} that depends on the separation between the
cylinders. This entropy differs from the usual polymer conformational entropy
and one should distinguish between the two. The identification of the wrapping
entropy presents a new concept in the analysis of the entropic effects in the context of polymer-particle
complexes. 

We finally calculate the force $\lambda$ required to sustain a given mean projected
separation $\langle d_\bot \rangle$ and derived the free energy $\Xi(l,\langle d_\bot \rangle)$.
For constrained wrapping we considered both symmetric and antisymmetric configurations with 
$\alpha_1 = \pm\alpha_2 = 3\pi/8, 5\pi/8$, respectively. In Fig. \ref{fig:interactions2}
we show $\Xi$ and $\lambda$ versus $\langle d_\bot \rangle/d_{max}$. It should be noted
that $\lambda \langle d_\bot \rangle > 0 (<0)$ corresponds to an intrinsic attractive 
(repulsive) force between the cylinders caused e.g. by charges on the cylinders. The behavior
of the $\lambda$ versus $\langle d_\bot \rangle/d_{max}$ curves is consistent with this interpretation.
We conclude that for the given choice of parameters the looped 
phase only occurs for symmetric wrapping and $\alpha_1 = \alpha_2 > \pi/2$. The range of $\lambda$   
chosen includes values where its magnitude exceeds the value of the applied external tension $f=0.4$. 
The response of $\langle d_\bot \rangle$ to $\lambda$ is as expected and we see a looped phase for a 
sufficiently repulsive intrinsic force.  Other
parameter choices can be investigated but we do not present the results here. In comparison, the case
of unconstrained wrapping is basically featureless and the results are shown in Figs. \ref{fig:interactionsC}
and \ref{fig:interactions3}, where only symmetric configuration is considered, the results for
antisymmetric configuration being very similar. We note that the force between cylinders is
repulsive in both these cases. We interpret this repulsion again as due to the wrapping entropy
that depends on the separation between the cylinders. Since the wrapping entropy might also play an 
important role in the case of multiple wrapped cylinders and could promote very strong non-pairwise additive
effects, we plan to study its effects very carefully in the future. Also we
intend to introduce a  {\em chemical potential}  for exchange of the wrapped cylinders
with a  {\em bulk} phase in order to generalize the calculation of the distribution
of nucleosomal core particles within the genomes \cite{Arneodo1,Arneodo2}.

A major conclusion of our work is that for constrained wrapping, where the amount of elastic filament 
wrapped around the cylinder subtends a fixed angle at the center, there are two kinds of transition that can occur as a 
function of the dimensionless external tension $f$, rigidity $\mu$ and wrapping energy $\sigma$. For two 
or more wrapped cylinders there is a transition from a 
looped to an extended phase which is additionally affected by the direct inter-cylinder forces, and there are 
desorption transitions which are sensitive to the symmetry of the wrapping (determined by the relationship
of the signs of wrapping angles) and also to the inter-cylinder separation.

A second major conclusion is that for unconstrained wrapping neither the looped phase nor the desorption
transition are likely to exist. Instead, there is an unwrapping transition where amount of filament wrapped
on each cylinders rapidly decreases as the external tension $f$ passes through a critical value. Correspondingly,
the inter-cylinder distances rapidly increase from small to near maximum values within a very small tension interval.  

\section{acknowledgments}

RP acknowledges support of ARRS through research program P1-0055 and research 
project J1-4297 as well as the DOE grant DE-SC0008176. AN acknowledges support from the Royal Society, the Royal 
Academy of Engineering, and the British Academy. We gratefully acknowledge support from Aspen Center for Physics, where
this work was initiated during the workshop on {\em New Perspectives in Strongly Correlated Electrostatics in Soft Matter} (2010). 
We would also like to thank Martin Muller for introducing us to his work on the interaction between colloidal cylinders on membranes, which 
provided the initial inspiration for this work. 

\appendix

\section{Constrained wrapping expressed in terms of the Mathieu functions}\label{appendix2}


The path integral $K$ obeys the Schr\" odinger equation
\begin{equation}
{\partial K(\psi,\psi',l)\over \partial l}=-HK,
\end{equation}
with boundary condition
\begin{equation}
K(\psi,\psi',0) = \delta(\psi-\psi').
\end{equation}
This clearly means that $K(\psi+2n\pi,\psi',l)\neq K(\psi,\psi',l)$ as it 
is violated at $l=0$ in the initial conditions. However, the propagator $K_M$ 
derived using the Mathieu functions, $\Psi_n$,
\begin{equation}
K_M(\psi,\psi',L) = \sum_n \exp(-E_nl)\Psi_n(\psi)\Psi_n(\psi') \end{equation}
has initial conditions
\begin{equation} 
K_M(\psi,\psi',0)=\sum_n \delta(\psi-\psi'-2n\pi),
\end{equation}
and is clearly periodic. We can thus write
\begin{equation}
K_M(\psi,\psi',l) = \sum_n  K(\psi,\psi'+2n\pi,l).
\end{equation}
If we take a single cylinder with fixed $\alpha$ (this is crucial in the argument that follows) we have 
\begin{widetext}\begin{equation}
Z(\alpha) = \int d\psi_0 d\psi_1  d \psi_3 ~K(\psi_0,\psi_1,l_1) S(\alpha,\psi_1)  
K(\psi_1 +\alpha,\psi_3, L-l_1 -R|\alpha |),
\end{equation}\end{widetext}
where $S$ is a general boundary terms which is periodic in $\psi_1$. The 
initial integral over $\psi_0$ can clearly be taken over the interval $[0,2\pi]$. If 
we restrict the integrals over $\psi_1$ and $\psi_2$ to $[0,2\pi]$ and 
add on their integer changes by hand we get
\begin{eqnarray}
Z(\alpha) &=& \int_0^{2\pi} d\psi_0 d\psi_1  d \psi_3 \sum_{m,n}
 K(\psi_0,\psi_1+2n\pi,l_1) S(\alpha,\psi_1+2n\pi) \times\nonumber\\
 &&\qquad\qquad\qquad\qquad\qquad\qquad \times\, K(\psi_1 +2n\pi +\alpha,\psi_3+2m\pi, L-l_1 -R|\alpha |),
\end{eqnarray}
where we explicitly show that we restrict the integrals to 
$[0,2\pi]$. Using the fact that $S$ is periodic for fixed $\alpha$ we finally obtain
\begin{equation}
Z(\alpha) = \int_0^{2\pi} d\psi_0 d\psi_1  d \psi_3 \sum_{m,n}K(\psi_0,\psi_1+2n\pi,l_1) 
S(\alpha_2,\psi_1) K(\psi_1 +2n\pi +\alpha,\psi_3+2m\pi, L-l_1 -R|\alpha |) ,
\end{equation}
and the obvious relation that
\begin{equation}
K(\psi+ 2n\pi,\psi'+2m\pi) = K(\psi,\psi'+2(m-n)\pi). 
\end{equation}
We then change the summation variable over $m$ to $m-n$ to obtain
\begin{eqnarray}
Z(\alpha)& =& \int_0^{2\pi} d\psi_0 d\psi_1  d \psi_3 \left(\sum_{n}K(\psi_0,\psi_1+2n\pi,l_1) 
\right)S(\alpha,\psi_1)\times\nonumber\\
&&\qquad\qquad\qquad\qquad\qquad\times\,\left(\sum_m K(\psi_1 +\alpha,\psi_3+2m\pi, L-l_1 -R|\alpha |)\right),
\end{eqnarray}
which then gives 
\begin{equation}
Z(\alpha) = \int_0^{2\pi} d\psi_0 d\psi_1  d \psi_3\, K_M(\psi_0,\psi_1) S(\alpha_2,\psi_1)
K_M(\psi_1 +\alpha,\psi_3, L-l_1 -R|\alpha |) ,
\end{equation}
which is the desired result expressed in terms of periodic Mathieu functions. The proof above for fixed wrapping angles can easily be extended to several cylinders with fixed wrapping angles. 

\section{Horizontal distance between the cylinders}\label{appendix1}

From Eq. (\ref{dublerw}) we can write for the horizontal distance between the cylinders
\begin{equation}
d_\bot = \la d_\bot^{l} \ra + R\,{\rm sgn}(\alpha_1)\la\sin(\psi_2)\ra-R\,{\rm sgn}(\alpha_2)\la\sin(\psi_3)\ra,
\end{equation}
where
\begin{widetext}
\begin{eqnarray}
\la d_\bot^{l} \ra &=&\frac{1}{Z(\alpha_1,\alpha_2,l)}\int ds \int d\psi_1d\psi_2\, \Psi_0(\psi_1)
 C_{\alpha_1}(\psi_1)\sum_{m,n}\exp({-\epsilon_m s/R}) \exp({-\epsilon_n (l-s)/R})\times \nn\\
&& \qquad\qquad\times \int  d(\sin{\psi_s})\Psi_m(\psi_s)\Psi_m(\psi_1+\alpha_1)\Psi_n(\psi_s) 
    \Psi_n(\psi_2)C_{\alpha_2}(\psi_2)\Psi_0(\psi_2+\alpha_2),
\end{eqnarray}
\end{widetext}
where a state has been inserted, with angle $\psi_s$ at the point $s$ of the 
first cylinder. We propagate the solution up to this point, calculate the 
horizontal projection, and then propagate to the remaining cylinder. We now find that 
\begin{widetext}
\begin{eqnarray}
\la d_\bot^{l} \ra&=&\frac{1}{Z(\alpha_1,\alpha_2,l)}\int d\psi_1d\psi_2\,  \Psi_0(\psi_1)\cdot
 C_{\alpha_1}(\psi_1)\sum_{m,n}\int ds\; \exp({-\epsilon_m s/R})\times \nn\\
&& \qquad\qquad\times\exp({-\epsilon_n (l-s)/R})D_{mn}(\psi_1+\alpha_1,\psi_2)
    C_{\alpha_2}(\psi_2)\Psi_0(\psi_2+\alpha_2),
\end{eqnarray}
where
\begin{equation}
D_{mn}(\psi_1+\alpha_1,\psi_2)=\int  d(\sin{\psi_s})\,P_m(\psi_1+\alpha_1,\psi_s)P_n(\psi_s,\psi_2), 
\end{equation}
and $C_{\alpha}$ is defined in Eq. (\ref{Calpha}).
By noting that after the angular integration only the exponential dependence on 
$s$ remains, we integrate $s$ over the range $0 \le s \le l$ to finally get
\begin{eqnarray}
\la d_\bot^{l} \ra&=&\frac{1}{Z(\alpha_1,\alpha_2,l)}\int d\psi_1d\psi_2\Psi_0(\psi_1)C_{\alpha_1}(\psi_1) 
\left[\sum_{n}le^{-\epsilon_n l/R}D_{nn}(\psi_1+\alpha_2,\psi_2)+ \right.\nn\\
& & \left. \qquad+\sum_{n,m\neq n}\frac{e^{{-\epsilon_m l/R}}-e^{{-\epsilon_n l/R}}}{\epsilon_n-\epsilon_m} 
D_{nm}(\psi_1+\alpha_2,\psi_2)\right] C_{\alpha_2}(\psi_2)\Psi_0(\psi_2+\alpha_2).
\end{eqnarray}
\end{widetext}
Note that because $L \gg l$, the effect of the filament lengths outside the 
cylinder region is encoded in the ground state factors $\Psi_0(\psi_1)$ and 
$\Psi_0(\psi_2+\alpha_2)$. The average distance $\la d_\bot \ra$ is easily 
computed since the various components can be pre-computed.
In a typical computation we pre-compute the lowest 20-100 Mathieu 
eigenfunctions and eigenvalues by recasting the Schr\"odinger equation in Eq. 
(\ref{sch_eqn}) as a matrix eigenvalue problem by discretizing the angle 
coordinate on the range $[0,2\pi]$, and using proprietory NAG routines 
\cite{NAG}. The computation of the partition function is then 
straightforward and can be done with modest computing resources.


\begin{thebibliography}{99}
\bibitem{comm} Note that as well as having a hard core interaction along the length of the filament there
is an additional hard core constraint coming from cylinders occupying different arc-lengths but sharing
the same position in space. Thus the simple Tonks gas model along the filament will contain additional long-range and non-local interactions.
\end{thebibliography}
\end{document}